\documentclass[aps,pra,showpacs,superscriptaddress,tightenlines,twocolumn,lengthcheck]{revtex4-2}
\usepackage{diagbox}
\usepackage{multirow}
\usepackage{amsmath}
\allowdisplaybreaks[4]  
\usepackage{amsfonts}
\usepackage{amsmath}
\usepackage{graphicx}
\usepackage{epsfig}
\usepackage{color}
\usepackage{txfonts}
\usepackage[colorlinks,citecolor=blue]{hyperref}
\usepackage{diagbox}
\usepackage{mathrsfs}
\usepackage{xcolor}
\usepackage{amsfonts}
\usepackage{bbm}
\usepackage{dsfont}
\usepackage[T1]{fontenc}
\usepackage{float}
\begin{document}
	
\title{Simultaneous ground-state cooling of six mechanical modes of two levitated nanoparticles}
		\author{Qian Zhang}
 \email{These authors contributed equally to this work.}
		\affiliation{Key Laboratory of Low-Dimensional Quantum Structures and Quantum Control of Ministry of Education, Key Laboratory for Matter Microstructure and Function of Hunan Province, Department of Physics and Synergetic Innovation Center for Quantum Effects and Applications, Hunan Normal University, Changsha 410081, China}
\affiliation{Hunan Research Center of the Basic Discipline for Quantum Effects and Quantum Technologies, Hunan Normal University, Changsha 410081, China}
		\author{Yi Xu}
 \email{These authors contributed equally to this work.}
		\affiliation{Key Laboratory of Low-Dimensional Quantum Structures and Quantum Control of Ministry of Education, Key Laboratory for Matter Microstructure and Function of Hunan Province, Department of Physics and Synergetic Innovation Center for Quantum Effects and Applications, Hunan Normal University, Changsha 410081, China}
\affiliation{Hunan Research Center of the Basic Discipline for Quantum Effects and Quantum Technologies, Hunan Normal University, Changsha 410081, China}
		\author{Jie-Qiao Liao}
		\email{Contact author: jqliao@hunnu.edu.cn}
		\affiliation{Key Laboratory of Low-Dimensional Quantum Structures and Quantum Control of Ministry of Education, Key Laboratory for Matter Microstructure and Function of Hunan Province, Department of Physics and Synergetic Innovation Center for Quantum Effects and Applications, Hunan Normal University, Changsha 410081, China}
        \affiliation{Hunan Research Center of the Basic Discipline for Quantum Effects and Quantum Technologies, Hunan Normal University, Changsha 410081, China}
		\affiliation{Institute of Interdisciplinary Studies, Hunan Normal University, Changsha 410081, China}
	
\begin{abstract}
Ground-state cooling is a prerequisite for exploring macroscopic quantum effects in mechanical motion of massive objects. Here we construct a polarization-angle-controllable coupled cavity-levitated-nanoparticle system in which two nanoparticles trapped by individual tweezers are coupled to a single-mode field in a cavity. We also study the simultaneous ground-state cooling of six mechanical displacement modes of the two levitated nanoparticles through the coherent scattering mechanism. By deriving the Hamiltonian of the system and performing the linearization, we obtain a linearized seven-mode Hamiltonian, which can exhibit the coupling structure and cooling mechanism. We confirm the physical condition for the appearance of dark modes, which will suppress the simultaneous ground-state cooling of these mechanical modes. We also find that, by properly tuning the polarization angle $\theta $ between the cavity field and the optical tweezer fields, the coupling channels can be controlled on demand and simultaneous ground-state cooling of these six motional modes of the two nanoparticles can be realized. Our work paves the way for generation and manipulation of collective macroscopic quantum effects in multiple levitated nanoparticles.
\end{abstract}

\date{\today }
\maketitle

\section{Introduction}
The optical tweezers~\cite{Ash1,Ash3,Ash4}, using the laser-beam-created optical gradient force to balance the scattering force and to achieve stable and non-contact trapping of particles, have become a fundamental technique for manipulating micro- and nanoscale particles. In recent years, with the development of laser and micro-nano manufacturing technologies, the optical tweezers have emerged as a premier physical system~\cite{JTR2020,CSC2021,Gaxv2307,MAarXiv2025} characterized by low noise, high isolation, and exceptional controllability. As a result, the optical tweezers have found widespread applications in physics~\cite{Ashkin1997,DJHPRL2009,KPG2010} and biology~\cite{AJTN1987,AKN1990,FSNP2011}, and established a viable platform for studying the fundamental of quantum theory such as macroscopic quantum phenomena and quantum-classical boundary~\cite{MADPRL2024,FARNPI2024}. Recent advances in this field include the realization of quantum squeezing in the motion of a levitated nanomechanical oscillator, enabled by rapid frequency modulation~\cite{MNKS2025}, and the demonstration of the delocalization of the quantum ground state of an optically levitated nanosphere by modulating the stiffness of the confining potential~\cite{MANAPRL2024}, highlighting the potential of levitated optomechanical systems for exploring macroscopic quantum effects.

In pursuit of superior means of manipulation and detection, the coupling of levitated nanoparticles (NPs) with optical cavities has recently emerged as a new research field: levitated optomechanics~\cite{DCS2010,PM2010,OMR2010}. The levitated optomechanical systems can achieve strong coupling between optical and mechanical modes~\cite{ANRNC2021,KJPRR2024,SZHAP2025}, providing a versatile platform for exploring quantum entanglement~\cite{HPRA2020,ANJP2020,IDTQS2021}, sensing~\cite{GMA2016,FSPRA2017,JJP2020,Zheng2020,FWA2020,TRWNP2023,TSPFR2023,ZLXPRR2023}, and information processing~\cite{HXZAQ2021,FAPNC2017}. Specifically, the primary condition for exploring macroscopic quantum effects in optomechanical systems is the cooling of the mechanical systems to their ground states~\cite{IPRL2007,FPRL2007,JTDN2011,JN2011,CPRL2019,CPRL2019,Magrini2021,Tebbenjohanns2021,LiuPRA2022}. In levitated optomechanics, the ground-state cooling of the center-of-mass motion of levitated particles via coherent scattering has recently been proposed~\cite{UPRL2019,CPRA2019,YYPRA2024}. By precisely controlling the frequency and position of the trapping laser relative to the cavity, ground-state cooling of a levitated nanoparticle along the $x$-direction was achieved with a final phonon number $n_{x}=0.43$~\cite{US2020}. Subsequently, simultaneous cooling along the $x$- and $y$-directions was realized via coherent scattering, with phonon numbers of $0.83$ and $0.81$, respectively~\cite{JNP2023}. More recently, three-dimensional (3D) cavity cooling of a single nanoparticle was demonstrated, with efficiency depending on its position in the cavity~\cite{DPRL2019}. In addition, the ground-state cooling of the librational mode of levitated particles via coherent scattering has also been demonstrated~\cite{LOJN2025,QAAN2025}. These advances demonstrate the extension of ground-state cooling from a single degree of freedom to multiple dimensions, laying a solid foundation for exploring quantum effects in more complex systems.

Motivated by the investigation of collective macroscopic quantum coherence in this platform, much recent attention has been paid to the manipulation of multiple particles and even complex geometric particle arrays~\cite{JZPRL,SA2020,JPR2023,Wu2025}. In particular, these systems are also widely applied in cutting-edge fields such as quantum precision measurement and quantum sensing~\cite{HPRL2022,YCJPRAP2023}. Recent experiments have achieved controllable coupling between two optically levitated nanoparticles via light-induced dipole-dipole interactions~\cite{JMS2022}. Furthermore, the realization of cavity-mediated long-range interactions has enabled scalable quantum control~\cite{JVarx2023}. In addition, the emergence of non-Hermitian physics and nonreciprocal coupling in two optically coupled levitated nanoparticles has enabled the observation of PT-symmetry breaking and self-sustained limit cycles~\cite{Varx2023,Marx2023}. Despite great advances have been achieved in manipulation of multi-particle systems, the simultaneous ground-state cooling of multiple particles remains a challenge for developing and utilizing quantum effects and technology in the multiple-particle platform.

To achieve this goal, we propose a polarization-angle-controllable coupled cavity-levitated-nanoparticle system to implement the simultaneous ground-state cooling of two particles in 3D displacement motions via coherent scattering. Based on the fact that the enhancement of coherent scattering by the optical cavity depends on the polarization direction of the optical tweezers, we introduce the polarization angle $\theta $ between the Fabry-P\'{e}rot cavity and the optical tweezers to control the coupling configuration and further the cooling performance. We find that the $Y$-axis motion decouples from other degrees of freedom at $\theta =0$ and $\pi $. Further, we demonstrate that tuning $\theta $ will effectively activate the coupling channels along the $Y$-axis, thereby enabling the simultaneous cooling of the particles in three dimensions. In particular, we find that the cooling efficiency depends sensitively on the polarization angle. By choosing proper work points during a wide range, the simultaneous ground-state cooling of these six mechanical modes of the two nanoparticles can be realized. This study provides a theoretical foundation for achieving 3D ground-state cooling in multiple-particle systems.

The rest of the work is organized as follows. In Sec.~\ref{PHYSICAL MODEL AND HAMILTONIANS}, we construct the theoretical model for the system of two levitated nanoparticles trapped in a Fabry-P\'{e}%
rot cavity, and present the Hamiltonian of the system. In Sec.~\ref{Linearization and covariance matrix}, we linearize the quantum Langevin equations around the steady state of the system and obtain the linearized seven-mode Hamiltonian. We also obtain the covariance matrix of the system. In Sec.~\ref{SIMULTANEOUS GROUND-STATE COOLING OF SIX MECHANICAL MODES}, we investigate the simultaneous ground-state cooling of the 3D motional degrees of freedom of the two particles. Finally, a brief discussion and summary are provided in Sec.~\ref{conclusion and discussions}. An Appendix is presented to show the detailed derivation of the interaction Hamiltonians.

\section{PHYSICAL SYSTEM AND HAMILTONIANS}\label{PHYSICAL MODEL AND HAMILTONIANS}

\begin{figure}[t!]
\center\includegraphics[width=0.48\textwidth]{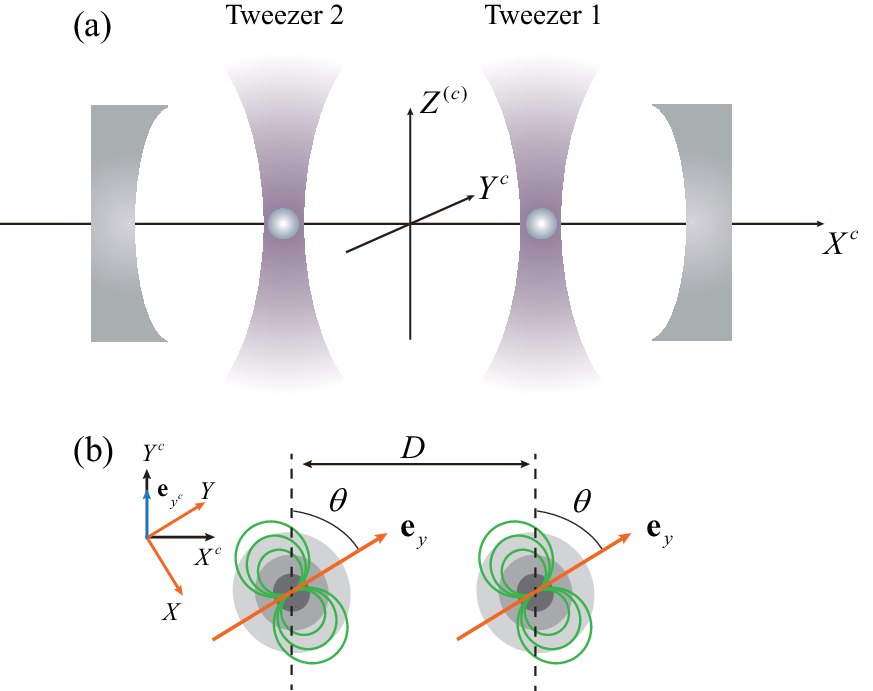}
\caption{{Schematic of the coupled cavity-levitated-nanoparticle system. (a) Side view: Two levitated NPs are trapped by optical tweezers and coupled to the field in a Fabry-P\'{e}rot cavity. The cavity axis is aligned along the $X^{c}$-direction, while the optical tweezers propagate along the $Z^{c}(Z)$-direction. (b) Top view: The cavity field plane $X^{c}-Y^{c}$ is rotated by an angle $\theta $ relative to the optical tweezer plane $X-Y$. The cavity polarization $\mathbf{e}_{y^{c}}$ (blue arrow) is along the along $Y^{c}$-axis, while the tweezer polarization $\mathbf{e}_{y}$ (orange arrow) is along the $Y$-axis. The gray ellipse represents the transverse optical tweezer potential, and the green ellipse corresponds to the dipole radiation field of the NPs. The distance between the two NPs is $D$.}}
\label{modelv1}
\end{figure}

We consider a coupled cavity-levitated-nanoparticle system, where two
NPs trapped by individual optical tweezers are coupled to a single-mode field in a Fabry-P\'{e}rot cavity, as shown in Fig.~\ref{modelv1}. The axis of the Fabry-P\'{e}rot cavity is along the $X^{c}$-direction, and the propagation of the optical tweezers is along the $Z^{c}$-axis. The two identical NPs are trapped at positions $\mathbf{\hat{R}}_{1}$ and $\mathbf{\hat{R}}_{2}$, with a spacing of $D$ along the cavity axis, and their parameters include: the radius $r_{0}=70$~nm, density $\rho \approx2200$ kg/m$^{3}$, and relative dielectric constant $\epsilon _{r}=2.07$.

The total Hamiltonian of the system reads
\begin{eqnarray}
\hat{H}_{\text{tot}} &=&\hat{H}_{\text{np}}+\hat{H}_{\text{cav}}+\hat{H}_{\text{int}}, \label{1}
\end{eqnarray}%
with
\begin{subequations} \label{2}
\begin{align}
\hat{H}_{\text{np}}&=\sum_{j=1,2}\frac{\mathbf{\hat{P}}_{j}^{2}}{2m_{j}}, \\
\hat{H}_{\text{cav}}&=\hbar \omega _{\text{cav}}\hat{a}^{\dag }\hat{a}, \\
\hat{H}_{\text{int}}&=-\frac{1}{2}\sum_{j=1,2}\alpha \mathbf{E}%
^{2}( \mathbf{\hat{R}}_{j}) \label{2c} .
\end{align}
\end{subequations}%
Here, $\hat{H}_{\text{np}}$ is the Hamiltonian corresponding to the kinetic energy of the two NPs, and $\mathbf{\hat{P}}_{j}$ and $m_{j}$ denote the momentum operator and mass of the $j$th $\left( j=1,2\right)$ NP, respectively. $\hat{H}_{\text{cav}}$ is the free Hamiltonian of the single-mode cavity field, and $\hat{a}$ $( \hat{a}^{\dag }) $ is the annihilation (creation) operator of the cavity mode. The third term $\hat{H}_{\text{int}}$ in Eq.~(\ref{1}) represents the interaction Hamiltonian between the NPs and the electric fields in the Rayleigh regime, in which the radius of the NPs is much smaller than the wavelength of the light, i.e., $ r_{0}\ll \lambda $~\cite{UPRL2019,DPRL2019,CPRA2019}.
In Eqs.~(\ref{2}), $\alpha =\varepsilon _{0}\epsilon_{c}V$ is the polarizability of the NP with $\epsilon _{c}=3( \epsilon _{r}-1) /(\epsilon _{r}+2) $, $\varepsilon _{0}$ the permittivity of free space, and $V=4\pi r_{0}^{3}/3$ the NP volume. $\mathbf{E}( \mathbf{\hat{R}}_{j})$ represents the total electric field at the position $\mathbf{\hat{R}}_{j}$ of the $j$th NP. The
center-of-mass position operator $\mathbf{\hat{R}}_{j}$ of the $j$th NP can be expressed as the sum of the focal coordinates  $\mathbf{r}_{j0}=(
x_{j0},y_{j0},0) $ of the $j$th optical tweezers and the position operator $\mathbf{\hat{r}}_{j}=( \hat{x}_{j},\hat{y}_{j},\hat{z}_{j})$ of the $j$th
NP, namely $\mathbf{\hat{R}}_{j}=\mathbf{r}_{j0}+\mathbf{\hat{r}}_{j}$. In this system, the total electric field $\mathbf{E}( \mathbf{\hat{R}}_{j})$ at the position $\mathbf{\hat{R}}_{j}$ of the $j$th NP is composed of three contributions: the cavity field $\hat{\mathbf{E}}_{\text{cav}}$, the optical tweezer field $\mathbf{E}_{\text{tw}}$, and the scattering field $\mathbf{E}_{G}$ induced by the other NP. In the following, we will introduce the expressions of these components.

For the cavity, we consider a single-mode cavity field along the $Y^{c}$-axis ($\mathbf{e}_{\text{cav}}=\mathbf{e}_{y^{c}}$), then the electric field $\hat{\mathbf{E}}_{\text{cav}}\left( \mathbf{r}\right) $ of the single cavity mode is given by%
\begin{equation}
\hat{\mathbf{E}}_{\text{cav}}\left( \mathbf{r}\right) =\epsilon _{\text{cav}}\cos \left(
k x^{c}-\phi \right) ( \hat{a}+\hat{a}^{\dag }) \mathbf{e}%
_{\text{cav}}, \label{5}
\end{equation}%
where $k=\omega _{\text{cav}}/c$ is the wave number, $\omega _{\text{cav}}$ is the cavity frequency, and $c$ is the speed of light in vacuum. $x^{c}$ is the distance of the cavity coordinate from the origin, and $\phi$ is the phase factor of the cavity field. In this work, we take $\phi = 0$ for simplicity. The field amplitude at the center of the cavity is given by $\epsilon _{\text{cav}}=\sqrt{%
\hbar \omega _{\text{cav}}/(2\varepsilon _{0}V_{\text{cav}})}$, with $V_{\text{cav}}$ denoting the cavity volume.

For the optical tweezer, we consider its fields as coherent Gaussian fields with a polarization direction along $\mathbf{e}_{\text{tw}}$. The optical tweezer field at point $\mathbf{r}$ can be written as~\cite{CPRA2019},%
\begin{equation}
\mathbf{E}_{\text{tw}}\left( \mathbf{r,}t\right) =\frac{1}{2}[ \mathbf{E}%
_{\text{tw}}\left( \mathbf{r}\right) e^{i\omega _{\text{tw}}t}+\mathbf{E}_{\text{tw}}^{\ast
}\left( \mathbf{r}\right) e^{-i\omega _{\text{tw}}t}] ,
\end{equation}%
with
\begin{equation}
\mathbf{E}_{\text{tw}}\left( \mathbf{r}\right) =\epsilon _{\text{tw}}\frac{W_{t}}{W\left(
z\right) }e^{-\frac{( x-x_{j0}) ^{2}}{W_{x}^{2}\left( z\right) }%
}e^{-\frac{( y-y_{j0}) ^{2}}{W_{y}^{2}\left( z\right) }%
}e^{ik_{\text{tw}}z}e^{i\phi _{t}\left( \mathbf{r}\right) }\mathbf{e}_{\text{tw}}. \label{3}
\end{equation}%
Here, $\omega _{\text{tw}}=ck_{\text{tw}}$ is the laser frequency and $k_{\text{tw}}=2\pi /\lambda _{\text{tw}}$ is the wave number of the laser, with $\lambda _{\text{tw}}=1064$~nm being the wavelength of the tweezer laser in vacuum. The beam waists along the $X$- and $Y$-axes are given by $W_{x,y} (z)=A_{x,y}W(z)$, with $A_{x,y}$ being dimensionless scaling factors. $W\left( z\right) =W_{t}\sqrt{1+\left( z/z_{R}\right) ^{2}}%
,$ where $W_{t}$ is the tweezer waist at the focus, and $z_{R}=\pi
W_{t}^{2}/\lambda _{\text{tw}}$ is the Rayleigh range. $\epsilon _{\text{tw}}=\sqrt{4P_{\text{tw}}/(\pi \varepsilon _{0}cW_{t}^{2}A_{x}A_{y})}$ is the amplitude of the electric field, with $P_{\text{tw}}$ being the power of the tweezer laser. In Eq.~(\ref{3}), the phase factor $\phi_{t}\left( \mathbf{r}\right) $ is given by%
\begin{equation}
\phi _{t}\left( \mathbf{r}\right) =\arctan \left( \frac{z}{z_{R}}\right) -%
\frac{k_{\text{tw}}z}{2}\frac{( x-x_{j0}) ^{2}+( y-y_{j0})
^{2}}{z^{2}+z_{R}^{2}} ,
\end{equation}%
where these variables have been introduced before. Since the Rayleigh range $z_{R}$ is typically several orders of magnitude larger than other relevant length scales in typical experiments~\cite{DPRL2019}. Throughout this work, we consider the case $\phi _{t}\left( \mathbf{r}\right)=0 $~\cite{JMS2022}.

To investigate the polarization-dependent coupling between the tweezers field and the cavity field, we introduce the angle $\theta \in \left[ 0,2\pi \right) $ between the polarization direction $\mathbf{e}_{\text{tw}}$ of the optical tweezer field and the polarization direction $\mathbf{e}_{y^{c}}$ of the cavity field. For calculation convenience, our analyses will be carried out in the ($X$-$Y$-$Z$) coordinate system, where the $Y$-axis is defined along the polarization direction $\mathbf{e}_{\text{tw}}$ of the trapping lasers, and the $Z$-axis is parallel to the $Z^{c}$-axis. The cavity $X^{c}$-$Y^{c}$ plane is rotated by an angle $\theta$ with respect to the tweezer $X$-$Y$ plane. When $\theta =0$, the polarization direction of the cavity field coincides with that of the optical tweezers. The coordinate transformation associated with the rotation of an angle $\theta $ is given by%
\begin{subequations} \label{6}
\begin{align}
X^{c} &= X\cos \theta +Y\sin \theta ,\\
Y^{c} &= -X\sin \theta +Y\cos \theta ,
\end{align}
\end{subequations}%
where $(X,Y)$ and $(X^{c},Y^{c})$ are the coordinates of a point in the two coordinate systems.

In this system, we only consider the physical processes of first-order scatterings since $\alpha$ is a small quantity~\cite{JMS2022,KPG2010}. Under the approximation, the induced radiation fields can be explicitly obtained. Concretely, $\mathbf{E}_{\text{Gtw}}^{\left( \bar{j}\right)
}( \mathbf{\hat{R}}_{j},t) =\overleftrightarrow{\mathbf{G}}%
_{\text{tw}}( \mathbf{\hat{R}}_{0}) \cdot \alpha \mathbf{E}%
_{\text{tw}}^{\left( \bar{j}\right) }( \mathbf{\hat{R}}_{\bar{j}},t) $
denotes the radiation field at $\mathbf{\hat{R}}_{\bar{j}}$\ generated by
the dipole induced by the $\bar{j}$th optical tweezer field, and $\mathbf{E}%
_{\text{Gcav}}( \mathbf{\hat{R}}_{j}) =\overleftrightarrow{\mathbf{G}}%
_{\text{cav}}( \mathbf{\hat{R}}_{0}) \cdot \alpha \mathbf{E}%
_{\text{cav}}( \mathbf{\hat{R}}_{\bar{j}}) $ represents the radiation
field from the dipole induced by the cavity field at $\mathbf{\hat{R}}_{\bar{%
j}}$. In these expressions, $\mathbf{\hat{R}}_{0}=( \hat{X}_{0},\hat{Y}_{0},\hat{Z}_{0})=\mathbf{\hat{R}}_{1}-\mathbf{\hat{R}}_{2}$, the index $\bar{j}$ denotes the other particle with respect to the $j$th particle (i.e., $\bar{1}=2$ and $\bar{2}=1$), and $\overleftrightarrow{\mathbf{G}}$ is the dyadic Green's function~\cite{KPG2010,Lbook2012},%
\begin{align}
\overleftrightarrow{\mathbf{G}}\left( \mathbf{R}_{0}\right)
= \frac{e^{ik_{0}R_{0}}}{4\pi \varepsilon _{0}R_{0}}
\begin{aligned}[t]
\Biggl[ & \left( \frac{1-ik_{0}R_{0}}{R_{0}^{2}}\right)
          \frac{3\mathbf{R}_{0}\mathbf{R}_{0}-R_{0}^{2}}{R_{0}^{2}} \\
       & + k_{0}^{2}\frac{R_{0}^{2}-\mathbf{R}_{0}\mathbf{R}_{0}}{R_{0}^{2}} \Biggr],
\end{aligned}
\end{align}
where $k_{0}$ is the wave number of the incident field. $\left\vert \mathbf{R}_{0} \right\vert =\left\vert \mathbf{R}_{1}-\mathbf{R}_{2} \right\vert$ is the distance between the two dipoles, and $\mathbf{R}_{0}\mathbf{R}_{0}=\sum_{\mu \mu^{\prime }}\mu_{0}\mu_{0}^{\prime }\mathbf{e}_{\mu}\mathbf{e}_{\mu^{\prime }}$, where $\mathbf{e}_{\mu}$ is the unit vector in the $\mu$ direction and $\mu,\mu^{\prime } \in \{x,y,z\}$. In addition, the subscript "cav" in the Green function $\overleftrightarrow{\mathbf{G}}_{\text{cav}}$ indicates that the incident field is the cavity field, i.e., $k_{0}=k$, while the subscript "tw" in $\overleftrightarrow{\mathbf{G}}_{\text{tw}}$ indicates that the incident field is the tweezer field, i.e., $k_{0}=k_{\text{tw}}$.

Based on the above analyses and Eqs.~(\ref{5}) and~(\ref{3}), the interaction Hamiltonian in Eq.~(\ref{2c}) can be rewritten as%
\begin{align}
\hat{H}_{\text{int}} = -\frac{1}{2}\alpha \sum_{j=1,2}
\Bigl[ & \mathbf{E}_{\text{tw}}^{\left( j\right)}( \mathbf{\hat{R}}_{j},t)
        + \mathbf{E}_{\text{cav}}( \mathbf{\hat{R}}_{j}) \notag \\
       & + \mathbf{E}_{\text{Gtw}}^{\left( \bar{j}\right) }( \mathbf{\hat{R}}_{j},t)
        + \mathbf{E}_{\text{Gcav}}( \mathbf{\hat{R}}_{j}) \Bigr]^{2}.
\label{7}
\end{align}%
Before expanding the square terms in Eq.~(\ref{7}), we present some analyses concerning the order of $\alpha$ for those terms in Eq.~(\ref{7}). The former two terms $\mathbf{E}_{\text{tw}}^{( j)}( \mathbf{\hat{R}}_{j},t)$ and $\mathbf{E}_{\text{cav}}( \mathbf{\hat{R}%
}_{j})$ in Eq.~(\ref{7}) are independent of $\alpha$, and the latter two terms $\mathbf{E}_{\text{Gtw}}^{( \bar{j}) }( \mathbf{\hat{R}}_{j},t)$ and $\mathbf{E}_{\text{Gcav}}( \mathbf{\hat{R}}_{j})$ should be first-order functions of $\alpha$. Therefore, to keep the terms up to the first order of $\alpha$ [apart from the factor $\alpha/2$ in Eq.~(\ref{7})], we need to discard the square term $[ \mathbf{E}_{\text{Gtw}}^{ (\bar{j}) }( \mathbf{\hat{R}}_{j},t) +\mathbf{E}_{\text{Gcav}}( \mathbf{\hat{R}}_{j}) ]^{2}$ during the expansion of Eq.~(\ref{7}). This is because this term describes the interaction between the radiation fields generated by the dipoles. In other words, this term corresponds to multiple-scattering processes and will be neglected in the following discussions. The details concerning the parameters in Eq.~(\ref{7}) will be given in the Appendix.

After a lengthy derivation, the total Hamiltonian of the system can be expressed in the rotating frame defined by the unitary transformation operator $\hat{U}=\exp(-i\omega_{\text{tw}} t \hat{a}^{\dagger}\hat{a})$ as
\begin{equation}
\begin{aligned}[b]
\hat{H}_{\text{tot}} &= \sum_{j=1,2}\frac{\mathbf{\hat{P}}_{j}^{2}}{2m_{j}}+ \sum_{\mu=x,y,z}\frac{1}{2}m\tilde{\omega}_{j\mu}^{2} \hat{\mu}_{j}^{2}  \\
&\quad +\hbar \tilde{R}_{x}\left( \hat{x}_{1}-\hat{x}_{2}\right) +\hbar \tilde{R}_{y}\left( \hat{y}_{1}-\hat{y}_{2}\right) \\
&\quad +\hbar \Delta ^{\prime }\hat{a}^{\dag }\hat{a}
+\hbar ( \tilde{\Omega}\hat{a}+\tilde{\Omega}^{\ast }\hat{a}^{\dag }) \\
&\quad +\sum_{j=1,2}\hbar [ \hat{a}( \tilde{g}_{x_{j}}\hat{x}_{j}+\tilde{g}_{y_{j}}\hat{y}_{j}+i\tilde{g}_{z_{j}}\hat{z}_{j}) + \text{H.c.}] \\
&\quad +\sum_{j=1,2}\hbar \tilde{g}_{ax_{j}}\hat{a}^{\dag }\hat{a}\hat{x}_{j}+\sum_{j=1,2}\hbar \tilde{g}_{ay_{j}}\hat{a}^{\dag }\hat{a}\hat{y}_{j} \\
&\quad -\sum_{\mu=x,y,z}k_{\mu}\hat{\mu}_{1}\hat{\mu}_{2}+\sum_{j=1,2}k_{xy}\hat{x}_{j}( \hat{y}_{j}-\hat{y}_{\bar{j}}) . \label{20}
\end{aligned}
\end{equation}%
Here, $\tilde{\omega}_{j\mu}=\sqrt{\omega_{j\mu}^{2}+2\nu_{\mu}/m+k_{\mu}/m}$ is the resonance frequency of the $\mu$-mode motion for the $j$th particle, with $\mu=x,y,z$ and $j=1,2$. $\tilde{R}_{x} = R_{\alpha} + g_{\alpha}/2$ and $\tilde{R}_{y} = R_{\beta} + g_{\beta}/2$ are the displacement magnitudes for the $X$- and $Y$-directional motions, respectively. The effective driving detuning is $\Delta ^{\prime }=\Delta + \sum_{j} \omega_{\text{sh}}^{\left( j\right) }-4\alpha \epsilon _{\text{cav}}^{2}\eta _{f}\cos ^{2}\theta
\cos \left( kD\right) \cos ^{2}\left( kD/2\right) /\hbar $ with $\Delta =\omega
_{\text{cav}}-\omega _{\text{tw}}$ and $\omega _{\text{sh}}^{\left( j\right) }=$\ $-2\alpha \epsilon
_{\text{cav}}^{2}\cos ^{2}\left( kD/2\right) /\hbar .$ $\tilde{\Omega} = \Omega^{\left( j\right)} + \Omega_{\alpha} + \Omega_{\beta}$ is the displacement magnitude for the cavity field. In Eq.~(\ref{20}), these optomechanical coupling strengths are introduced by
\begin{subequations}
\begin{align}
\tilde{g}_{\mu_{j}} &= g_{\mu_{j}} + g_{\alpha \mu_{j}} + g_{\beta \mu_{j}}, \\
\tilde{g}_{ax_{j}} &= g_{ax_{j}} - ( -1) ^{j} g_{\alpha}, \\
\tilde{g}_{ay_{j}} &= g_{ay_{j}} - ( -1) ^{j} g_{\beta}.
\end{align}
\end{subequations}%
Here, these variables have been defined in Eqs.~(\ref{18}), ~(\ref{A6}), ~(\ref{A8}), and ~(\ref{A10}).

\section{Linearization and covariance matrix}\label{Linearization and covariance matrix}
The physical system under consideration is nonlinear, as shown by the Hamiltonian in Eq.~(\ref{20}). Concretely, there exist both bilinear and trilinear couplings between the cavity field and the $X$-, $Y$-direction mechanical motional modes, while only bilinear coupling exist between the cavity mode and the $Z$-direction mechanical motion modes. For the mechanical mode interactions, the $X$- and $Y$-direction mechanical motional modes will be directly mixed, but the $Z$-direction motional modes of the two particles are only coupled with each other. In addition, the cavity field is driven by a monochromatic field [described by the term $( \tilde{\Omega}\hat{a}+\tilde{\Omega}^{\ast }\hat{a}^{\dag })$], and the $X$- and $Y$-direction motions are displaced [described by $\tilde{R}_{x}$ and $\tilde{R}_{y}$ terms in Eq.~(\ref{20})]. Physically, in the strong driving and displacement regime, we can linearize the system around the steady semi-classical motion.

To perform the linearization, we first derive the quantum Langevin equations of the system. For convenience, we introduce the dimensionless position and momentum operators
\begin{equation}
\hat{q}_{\mu_{j}}=\hat{\mu}_{j}/\sqrt{2}\mu_{j,\text{zpf}}\text{,}\hspace{0.5cm}\hat{p}_{\mu_{j}}=\hat{P}_{\mu_{j}}/\sqrt{2}%
p_{\mu_{j},\text{zpf}} ,
\end{equation}
where $\mu_{j,\text{zpf}}=\sqrt{\hbar /(2m \tilde{\omega}_{j\mu})}$ and $p_{\mu_{j},\text{zpf}}=\sqrt{m \tilde{\omega}_{j\mu}\hbar /2}$ ($j=1,2$) are the zero-point motions associated with the coordinates and momentum, respectively. $\hat{P}_{\mu_{j}}$ is the $\mu$-direction momentum operator for the $j$th NP. Based on Eq.~(\ref{20}), the quantum Langevin equations describing the evolution of this system can be obtained as%
\begin{subequations}  \label{28}
\begin{align}
\dot{\hat{q}}_{\mu_{j}} &= \tilde{\omega}_{j\mu}\hat{p}_{\mu_{j}}, \label{24a} \\
\dot{\hat{p}}_{x_{1}} &= -\tilde{\omega}_{1x}\hat{q}_{x_{1}}-\tilde{R}_{x_{1}}-G_{ax_{1}}\hat{a}^{\dag}\hat{a}-G_{x_{1}}\hat{a}-G_{x_{1}}^{\ast}\hat{a}^{\dag} \notag \\
        &\quad +G_{x}\hat{q}_{x_{2}}-G_{x_{1}y_{1}}\hat{q}_{y_{1}}+G_{x_{1}y_{2}}\hat{q}_{y_{2}}-\gamma_{x_{1}}\hat{p}_{x_{1}}+\hat{f}_{x_{1}}^{th}, \\
\dot{\hat{p}}_{x_{2}} &= -\tilde{\omega}_{2x}\hat{q}_{x_{2}}+\tilde{R}_{x_{2}}-G_{ax_{2}}\hat{a}^{\dag}\hat{a}-G_{x_{2}}\hat{a}-G_{x_{2}}^{\ast}\hat{a}^{\dag} \notag \\
        &\quad +G_{x}\hat{q}_{x_{1}}-G_{x_{2}y_{2}}\hat{q}_{y_{2}}+G_{x_{2}y_{1}}\hat{q}_{y_{1}}-\gamma_{x_{2}}\hat{p}_{x_{2}}+\hat{f}_{x_{2}}^{th}, \\
\dot{\hat{p}}_{y_{1}} &= -\tilde{\omega}_{1y}\hat{q}_{y_{1}}-\tilde{R}_{y_{1}}-G_{ay_{1}}\hat{a}^{\dag}\hat{a}-G_{y_{1}}\hat{a}-G_{y_{1}}^{\ast}\hat{a}^{\dag} \notag \\
        &\quad +G_{y}\hat{q}_{y_{2}}-G_{x_{1}y_{1}}\hat{q}_{x_{1}}+G_{x_{2}y_{1}}\hat{q}_{x_{2}}-\gamma_{y_{1}}\hat{p}_{y_{1}}+\hat{f}_{y_{1}}^{th}, \\
\dot{\hat{p}}_{y_{2}} &= -\tilde{\omega}_{2y}\hat{q}_{y_{2}}+\tilde{R}_{y_{2}}-G_{ay_{2}}\hat{a}^{\dag}\hat{a}-G_{y_{2}}\hat{a}-G_{y_{2}}^{\ast}\hat{a}^{\dag} \notag \\
        &\quad +G_{y}\hat{q}_{y_{1}}-G_{x_{2}y_{2}}\hat{q}_{x_{2}}+G_{x_{1}y_{2}}\hat{q}_{x_{1}}-\gamma_{y_{2}}\hat{p}_{y_{2}}+\hat{f}_{y_{2}}^{th}, \\
\dot{\hat{p}}_{z_{1}} &= -\tilde{\omega}_{1z}\hat{q}_{z_{1}}-i G_{z_{1}}\hat{a}+iG_{z_{1}}^{\ast}\hat{a}^{\dag}+G_{z}\hat{q}_{z_{2}}-\gamma_{z_{1}}\hat{p}_{z_{1}}+\hat{f}_{z_{1}}^{th}, \\
\dot{\hat{p}}_{z_{2}} &= -\tilde{\omega}_{2z}\hat{q}_{z_{2}}-i G_{z_{2}}\hat{a}+i G_{z_{2}}^{\ast}\hat{a}^{\dag}+G_{z}\hat{q}_{z_{1}}-\gamma_{z_{2}}\hat{p}_{z_{2}}+\hat{f}_{z_{2}}^{th}, \\
\dot{\hat{a}} &= \left( -i\Delta ^{\prime }-\kappa \right) \hat{a}-i\tilde{\Omega}^{\ast} -iG_{ax_{1}}\hat{a}\hat{q}_{x_{1}}-iG_{ax_{2}}\hat{a}\hat{q}_{x_{2}} \notag \\
        &\quad -iG_{ay_{1}}\hat{a}\hat{q}_{y_{1}}-iG_{ay_{2}}\hat{a}\hat{q}_{y_{2}}-iG_{x_{1}}^{\ast}\hat{q}_{x_{1}}-iG_{x_{2}}^{\ast}\hat{q}_{x_{2}} \notag \\
        &\quad -iG_{y_{1}}^{\ast}\hat{q}_{y_{1}}-iG_{y_{2}}^{\ast}\hat{q}_{y_{2}}-G_{z_{1}}^{\ast}\hat{q}_{z_{1}}-G_{z_{2}}^{\ast}\hat{q}_{z_{2}}+\sqrt{2\kappa}\hat{a}_{in}\left(t\right),  \label{24m} \\
\dot{\hat{a}}^{\dag} &= \left( i\Delta ^{\prime }-\kappa \right) \hat{a}^{\dag}+i\tilde{\Omega}+iG_{ax_{1}}\hat{a}^{\dag}\hat{q}_{x_{1}}+iG_{ax_{2}}\hat{a}^{\dag}\hat{q}_{x_{2}} \notag \\
        &\quad +iG_{ay_{1}}\hat{a}^{\dag}\hat{q}_{y_{1}}+iG_{ay_{2}}\hat{a}^{\dag}q_{y_{2}}+iG_{x_{1}}\hat{q}_{x_{1}}+iG_{x_{2}}\hat{q}_{x_{2}} \notag \\
        &\quad +iG_{y_{1}}\hat{q}_{y_{1}}+iG_{y_{2}}\hat{q}_{y_{2}}-G_{z_{1}}\hat{q}_{z_{1}}-G_{z_{2}}\hat{q}_{z_{2}}+\sqrt{2\kappa}\hat{a}_{in}^{\dag}\left(t\right),
\end{align}
\end{subequations}%
where $\kappa $ and $\gamma _{\mu_{j}}$ are the decay rates of the cavity mode
and the $\mu$-direction motion of the $j$th NP with $\mu=x,y,z$ and $j=1,2$, respectively.

In Eqs.~(\ref{28}), $G_{ax_{j}}=\sqrt{2}\tilde{g}%
_{ax_{j}}x_{j,\text{zpf}},$ $G_{ay_{j}}=\sqrt{2}\tilde{g}_{ay_{j}}y_{j,\text{zpf}},$ $%
G_{\mu_{j}}=\sqrt{2}\tilde{g}_{\mu_{j}}q_{j,\text{zpf}},$ $G%
_{\mu_{j}}^{\ast }=\sqrt{2}\tilde{g}_{\mu_{j}}^{\ast }q_{j,\text{zpf}},$ $%
G_{\mu}=2k_{\mu}q_{1,\text{zpf}}q_{2,\text{zpf}}/\hbar ,$ and $%
G_{x_{j}y_{j}}=2k_{xy}x_{j,\text{zpf}}y_{j,\text{zpf}}/\hbar $ are the coupling
coefficients; $\tilde{R}_{x_{j}}=\sqrt{2}\tilde{R}_{x}x_{j,\text{zpf}}$ and $\tilde{R}_{y_{j}}=\sqrt{2}\tilde{R}_{y}y_{j,\text{zpf}}$ are the displacement magnitudes of the $x$ and $y$ modes; $\hat{f}_{\mu_{j}}^{th}$ is the stochastic thermal noise operator corresponding to the $\mu$ mode motion of the $j$th particle, which is determined by the zero average values $\langle \hat{f}_{\mu_{j}}^{th}\left( t\right)
\rangle =0$ and the correlation function%
\begin{eqnarray}
\langle \hat{f}_{\mu_{j}}^{th}\left( t\right) \hat{f}_{\mu_{j^{\prime }}^{\prime
}}^{th}\left( t^{\prime }\right) \rangle  &=&\delta _{jj^{\prime }}\delta
_{\mu \mu^{\prime }}\frac{\gamma _{\mu_{j}}}{\tilde{\omega}_{j\mu}}\int e^{-i\omega
\left( t-t^{\prime }\right) }\omega   \nonumber \\
&&\times \left[ \coth \left( \frac{\hbar \omega }{2k_{B}T_{\mu_{j}}}\right) +1%
\right] \frac{d\omega }{2\pi }.
\end{eqnarray}
Here, $k_{B}$ is the Boltzman constant, and $T_{\mu_{j}}$ is the temperature of the thermal bath associated with the $\mu_{j}$ mode. In the high-temperature limit, we can approximately get $\coth [\hbar \tilde{\omega}_{j\mu}/(k_{B}T_{\mu_{j}})] +1\approx
2k_{B}T_{\mu_{j}}/(\hbar \tilde{\omega}_{j\mu})$, then the symmetric correlation function of the stochastic noise can be obtained as%
\begin{eqnarray}
&&\langle \hat{f}_{\mu_{j}}^{th}\left( t\right) \hat{f}_{\mu_{j^{\prime }}^{\prime }}^{th}\left( t^{\prime }\right) +\hat{f}_{\mu_{j^{\prime }}^{\prime }}^{th}\left( t^{\prime
}\right) \hat{f}_{\mu_{j}}^{th}\left( t\right) \rangle   \nonumber \\
&\approx &2\gamma _{\mu_{j}}( 2\bar{n}_{\mu_{j}}^{th}+1) \delta (
t-t^{\prime }) \delta _{j j^{\prime }} \delta _{\mu \mu^{\prime }},
\end{eqnarray}%
where $\bar{n}_{\mu_{j}}^{th}=[\exp ( \hbar \tilde{\omega}%
_{j\mu}/k_{B}T_{\mu_{j}}) -1]^{-1}\approx k_{B}T_{\mu_{j}}/(\hbar \tilde{\omega}%
_{j\mu})$ is the thermal occupation number for the bath of the $\mu_{j}$ mode motion. In Eq.~(\ref{24m}), $\hat{a}_{in}$ ($\hat{a}_{in}^{\dag}$) is the noise operator related to the cavity mode $a$.

To study the optomechanical cooling, we prefer to work in the quadrature operator representation of the system. Therefore, we introduce the quadrature operators $\hat{X}_{a}=( \hat{a}^{\dag }+ \hat{a}) /\sqrt{2}$ and $ \hat{Y}_{a}=i( \hat{a}^{\dag }- \hat{a}) /\sqrt{2}$, as well as the corresponding input noise operators $\hat{X}_{in}=(
\hat{a}_{in}^{\dag }+\hat{a}_{in}) /\sqrt{2}$ and $\hat{Y}_{in}=i( \hat{a}_{in}^{\dag
}-\hat{a}_{in}) /\sqrt{2}.$ These optical noise operators are determined by the zero average values $\langle \hat{X}_{in}\rangle =0$ and $\langle \hat{Y}_{in}\rangle =0
$, and the correlation functions~\cite{Cbook2000}%
\begin{subequations}
\begin{align}
\langle \hat{X}_{in}\left( t\right) \hat{X}_{in}\left( t^{\prime }\right)
\rangle &= \frac{1}{2}\delta \left( t-t^{\prime }\right) , \\
\langle \hat{Y}_{in}\left( t\right) \hat{Y}_{in}\left( t^{\prime }\right)
\rangle &= \frac{1}{2}\delta \left( t-t^{\prime }\right) , \\
\langle \hat{X}_{in}\left( t\right) \hat{Y}_{in}\left( t^{\prime }\right)
\rangle &= \frac{i}{2}\delta \left( t-t^{\prime }\right) , \\
\langle \hat{Y}_{in}\left( t\right) \hat{X}_{in}\left( t^{\prime }\right)
\rangle &= -\frac{i}{2}\delta \left( t-t^{\prime }\right) .
\end{align}
\end{subequations}

Under the strong-driving condition, we can perform the linearization of the system by expressing the system operators as a summation of the steady-state average values and quantum fluctuation, $\hat{o}=\left\langle \hat{o}\right\rangle_{ss}+\delta \hat{o}$ for $\hat{o}=\hat{q}_{x_{1,2}}, \hat{q}_{y_{1,2}}, \hat{q}_{z_{1,2}}, \hat{X}_{a}$, and $\hat{Y}_{a}$. Then, the equations of motion for quantum fluctuations can be expressed as a compact matrix form
\begin{equation}
\mathbf{\dot{\hat{u}}}\left( t\right) =\mathbf{A\hat{u}}\left( t\right) +\mathbf{\hat{N}}%
\left( t\right), \label{30}
\end{equation}%
where we introduce the quadrature fluctuation operator vector
\begin{eqnarray}
\mathbf{\hat{u}}\left( t\right)  &=&(\delta \hat{X}_{a},\delta \hat{Y}_{a},\delta \hat{p}_{x_{1}},\delta \hat{p}_{x_{2}},\delta \hat{p}_{y_{1}},\delta \hat{p}_{y_{2}},\delta \hat{p}_{z_{1}},  \delta \hat{p}_{z_{2}}, \notag \\
&&\delta \hat{q}_{x_{1}},\delta \hat{q}_{x_{2}},\delta \hat{q}_{y_{1}},\delta \hat{q}_{y_{2}},\delta \hat{q}_{z_{1}},\delta \hat{q}_{z_{2}})^{\text{T}} ,
\end{eqnarray}%
with "T" denoting the matrix transpose. In Eq.~(\ref{30}), we also introduce the corresponding input noise operator vector
\begin{eqnarray}
\mathbf{\hat{N}}\left( t\right)  &=&(\sqrt{2\kappa }\delta \hat{X}_{in},\sqrt{2\kappa }\delta \hat{Y}_{in},\hat{f}_{x_{1}}^{th},\hat{f}_{x_{2}}^{th},\hat{f}_{y_{1}}^{th}, \hat{f}_{y_{2}}^{th},\hat{f}_{z_{1}}^{th},\hat{f}_{y_{2}}^{th},\notag \\
&& 0,0,0,0,0,0)^{\text{T}}.
\end{eqnarray}%
In addition, the coefficient matrix in Eq.~(\ref{30}) is introduced as
\begin{equation} \label{31}
\mathbf{A}=%
\begin{pmatrix}
\mathbf{A}_{p} & \mathbf{A}_{q} \\
\mathbf{0} & \mathbf{A}_{\omega}%
\end{pmatrix} ,
\end{equation}%
where these three block matrices in Eq.~(\ref{31}) are given by
\begin{subequations}
\begin{align}
\mathbf{A}_{p}&=%
\begin{pmatrix}
-\kappa & \tilde{\Delta} & 0 & 0 & 0 & 0 & 0 & 0 \\
-\tilde{\Delta} & \kappa & 0 & 0 & 0 & 0 & 0 & 0  \\
-2B_{1} & -2A_{1} & -\gamma _{x_{1}} & 0 & 0 & 0 & 0 & 0 \\
-2B_{2} & -2A_{2} & 0 & -\gamma _{x_{2}} & 0 & 0 & 0 & 0 \\
-2D_{1} & -2C_{1} & 0 & 0 & -\gamma _{y_{1}} & 0 & 0 & 0 \\
-2D_{2} & -2C_{2} & 0 & 0 & 0 & -\gamma _{y_{2}} & 0 & 0 \\
2E_{1} & 2F_{1} & 0 & 0 & 0 & 0 & -\gamma _{z_{1}} & 0 \\
2E_{2} & 2F_{2} & 0 & 0 & 0 & 0 & 0 & -\gamma _{z_{2}}%
\end{pmatrix},\\
\mathbf{A}_{q}&=%
\begin{pmatrix}
2A_{1} & 2A_{2} & 2C_{1} & 2C_{2} & 2F_{1} & -2F_{2} \\
-2B_{1} & -2B_{2} & -2D_{1} & -2D_{2} & 2E_{1} & 2E_{2}  \\
-\tilde{\omega}_{1x} & G_{x} & -G_{x_{1}y_{1}} & G_{x_{1}y_{2}} & 0 & 0  \\
G_{x} & -\tilde{\omega}_{2x} & G_{x_{2}y_{1}} & -G_{x_{2}y_{2}} & 0 & 0 \\
-G_{x_{1}y_{1}} & G_{x_{2}y_{1}} & -\tilde{\omega}_{1y} & G_{y} & 0 & 0 \\
G_{x_{1}y_{2}} & -G_{x_{2}y_{2}} & G_{y} & -\tilde{\omega}_{2y} & 0 & 0 \\
0 & 0 & 0 & 0 & -\tilde{\omega}_{1z} & G_{z}  \\
 0 & 0 & 0 & 0 & G_{z} & -\tilde{\omega}_{2z}%
\end{pmatrix},
\end{align}
\end{subequations}%
and
\begin{eqnarray}
\mathbf{A}_{\omega} & = & \text{diag}[\tilde{\omega}_{1x} ,\tilde{\omega}_{2x},\tilde{\omega}_{1y},
\tilde{\omega}_{2y},\tilde{\omega}_{1z} ,\tilde{\omega}_{2z} ].
\end{eqnarray}%
Here, $G_{\mu_{j}}^{\prime }=\sqrt{2} G_{\mu_{j}}/2=\tilde{g}%
_{\mu_{j}}q_{j,\text{zpf}},$ $G_{\mu_{j}}^{\prime \ast }=\tilde{g}%
_{\mu_{j}}^{\ast }q_{j,\text{zpf}}$, $G_{ax_{j}}^{\prime }=\tilde{g}_{ax_{j}}x_{j,\text{zpf}},$ $%
G_{ay_{j}}^{\prime }=\tilde{g}_{ay_{j}}y_{j,\text{zpf}},$ and $\tilde{\Delta}=\Delta ^{\prime
}+iG_{ax_{1}}\left\langle \hat{x}_{1}\right\rangle _{ss}+iG_{ax_{2}}\left\langle
\hat{x}_{2}\right\rangle _{ss}+iG_{ay_{1}}\left\langle \hat{y}_{1}\right\rangle_{ss}
+iG_{ay_{2}}\left\langle \hat{y}_{2}\right\rangle_{ss} $. We introduce $A_{j}=\mathrm{Im}[G_{ax_{j}}^{\prime }\left\langle \hat{a}%
\right\rangle _{ss}-G_{x_{j}}^{\prime }],$ $B_{j}=\mathrm{Re}%
[G_{ax_{j}}^{\prime }\left\langle \hat{a}\right\rangle_{ss} +G%
_{x_{j}}^{\prime }],$ $C_{j}=\mathrm{Im}[G_{ay_{j}}^{\prime }-G%
_{y_{j}}^{\prime }],$ $D_{j}=\mathrm{Re}[G_{ay_{j}}^{\prime }+G%
_{y_{j}}^{\prime }],$ $E_{j}=\mathrm{Im}[ G_{z_{j}}^{\prime }%
] ,$ and $F_{j}=\mathrm{Re}[ G_{z_{j}}^{\prime }].$
The expectation values of these operators are governed by the equations of motion for the following semiclassical motion,%
\begin{subequations}\label{21}
\begin{align}
&( i\tilde{\Delta}-\kappa ) \langle \hat{a}^{\dag }\rangle_{ss}+iG%
_{x_{1}}\left\langle \hat{x}_{1}\right\rangle_{ss}+iG_{x_{2}}\left\langle \hat{x}_{2}\right\rangle_{ss} +iG_{y_{1}}\left\langle
\hat{y}_{1}\right\rangle_{ss} \notag \\
&\quad  +iG_{y_{2}}\left\langle
\hat{y}_{2}\right\rangle _{ss}-G_{z_{1}}\left\langle \hat{z}_{1}\right\rangle_{ss}
-G_{z_{2}}\left\langle \hat{z}_{2}\right\rangle_{ss} +i\tilde{\Omega}=0, \\
&( -i\tilde{\Delta}-\kappa ) \left\langle \hat{a} \right\rangle_{ss} -iG%
_{x_{1}}^{\ast }\left\langle \hat{x}_{1}\right\rangle_{ss} -iG_{x_{2}}^{\ast
}\left\langle \hat{x}_{2}\right\rangle_{ss} -iG_{y_{1}}^{\ast }\left\langle
\hat{y}_{1}\right\rangle_{ss} \notag \\
&\quad  -iG_{y_{2}}^{\ast }\left\langle
\hat{y}_{2}\right\rangle _{ss}-G_{z_{1}}^{\ast }\left\langle \hat{z}_{1}\right\rangle_{ss}
-G_{z_{2}}^{\ast }\left\langle \hat{z}_{2}\right\rangle_{ss} -i\tilde{\Omega}%
^{\ast }=0, \\
&-\tilde{\omega}_{1x}\left\langle \hat{x}_{1}\right\rangle_{ss} -G_{ax_{1}}\langle
\hat{a}^{\dag }\rangle_{ss} \left\langle \hat{a}\right\rangle_{ss} -G%
_{x_{1}}\left\langle \hat{a}\right\rangle_{ss} -G_{x_{1}}^{\ast }\langle
\hat{a}^{\dag }\rangle_{ss} \notag \\
        &\quad  +G_{x}\left\langle \hat{x}_{2}\right\rangle_{ss}
-G_{x_{1}y_{1}}\left\langle \hat{y}_{1}\right\rangle_{ss} +G_{x_{1}y_{2}}\left\langle
\hat{y}_{2}\right\rangle_{ss} -\tilde{R}_{x_{1}}=0, \\
&-\tilde{\omega}_{2x}\left\langle \hat{x}_{2}\right\rangle_{ss} -G_{ax_{2}}\langle
\hat{a}^{\dag }\rangle_{ss} \left\langle \hat{a}\right\rangle_{ss} -G%
_{x_{2}}\left\langle \hat{a}\right\rangle_{ss} -G_{x_{2}}^{\ast }\langle
\hat{a}^{\dag }\rangle_{ss} \notag \\
        &\quad  +G_{x}\left\langle \hat{x}_{1}\right\rangle_{ss}
-G_{x_{2}y_{2}}\left\langle \hat{y}_{2}\right\rangle_{ss} +G_{x_{2}y_{1}}\left\langle
\hat{y}_{1}\right\rangle_{ss}  +\tilde{R}_{x_{2}}=0, \\
& -\tilde{\omega}_{1y}\left\langle \hat{y}_{1}\right\rangle_{ss} -G_{ay_{1}}\langle
\hat{a}^{\dag }\rangle_{ss} \left\langle \hat{a}\right\rangle_{ss} -G%
_{y_{1}}\left\langle \hat{a}\right\rangle_{ss} -G_{y_{1}}^{\ast }\langle
\hat{a}^{\dag }\rangle_{ss} \notag \\
        &\quad  +G_{y}\left\langle \hat{y}_{2}\right\rangle_{ss}
-G_{x_{1}y_{1}}\left\langle \hat{x}_{1}\right\rangle_{ss} +G_{x_{2}y_{1}}\left\langle
\hat{x}_{2}\right\rangle_{ss} -\tilde{R}_{y_{1}}=0, \\
& -\tilde{\omega}_{2y}\left\langle \hat{y}_{2}\right\rangle_{ss} -G_{ay_{2}}\langle
\hat{a}^{\dag }\rangle_{ss} \left\langle \hat{a}\right\rangle_{ss} -G%
_{y_{2}}\left\langle \hat{a}\right\rangle_{ss} -G_{y_{2}}^{\ast }\langle
\hat{a}^{\dag }\rangle_{ss} \notag \\
        &\quad  +G_{y}\left\langle \hat{y}_{1}\right\rangle_{ss}
-G_{x_{2}y_{2}}\left\langle \hat{x}_{2}\right\rangle_{ss} +G_{x_{1}y_{2}}\left\langle
\hat{x}_{1}\right\rangle_{ss} +\tilde{R}_{y_{2}}=0, \\
& -\tilde{\omega}_{1z}\left\langle \hat{z}_{1}\right\rangle_{ss} -iG%
_{z_{1}}\left\langle \hat{a}\right\rangle_{ss} +iG_{z_{1}}^{\ast }\langle
\hat{a}^{\dag }\rangle_{ss} +G_{z}\left\langle \hat{z}_{2}\right\rangle _{ss} =0, \\
& -\tilde{\omega}_{2z}\left\langle \hat{z}_{2}\right\rangle_{ss} -iG%
_{z_{2}}\left\langle \hat{a}\right\rangle_{ss} +iG_{z_{2}}^{\ast }\langle
\hat{a}^{\dag }\rangle_{ss} +G_{z}\left\langle \hat{z}_{1}\right\rangle_{ss} =0.
\end{align}
\end{subequations}%
Equations ~(\ref{21}) are a system of algebraic equations for the steady-state average values of the system, which contains the expectation value $\langle \hat{a} \rangle_{ss}$ of the optical field operator and the expectation values $\langle \hat{\mu}_{j}\rangle_{ss}$ of the mechanical displacements. From Eq.~(\ref{24a}), it can be seen that the expectation values of the momentum operators of each mechanical mode satisfy $\langle \dot{\hat{q}}_{\mu _{j}}\rangle_{ss} =\tilde{\omega}_{j\mu}\langle \hat{p}_{\mu_{j}}\rangle_{ss} =0$. Equation~(\ref{30}) describes the linear dynamics of the system, and hence its stability condition can be analyzed using the Routh-Hurwitz criterion~\cite{Ibook2014}. In our following simulations, all the parameters used satisfy the stability conditions.

The formal solution of the linearized Langevin equation~(\ref{30}) can be written as
\begin{equation}
\mathbf{\hat{u}}\left(
t\right) =\mathbf{M}\left( t\right) \mathbf{\hat{u}}\left( 0\right) +\int_{0}^{t}%
\mathbf{M}\left( t-s\right) \mathbf{\hat{N}}\left( s\right) ds  ,
\end{equation}%
where we introduce the matrix $\mathbf{M}\left( t\right) =\exp \left( \mathbf{A}t\right) $. The final mean phonon numbers of these six mechanical modes can be calculated by solving the steady state of the system. To this end, we introduce the covarince matrix $\mathbf{V}$, which is defined by the matrix elements%
\begin{equation}
\mathbf{V}_{n,m}\left( \infty \right) =\frac{1}{2} \left[ \left\langle \mathbf{\hat{u}}%
_{n}\left( \infty \right) \mathbf{\hat{u}}_{m}\left( \infty \right) \right\rangle
+\left\langle \mathbf{\hat{u}}_{m}\left( \infty \right) \mathbf{\hat{u}}_{n}\left(
\infty \right) \right\rangle \right]
\end{equation}%
for $n, m=1-14$. The covariance matrix is determined by the Lyapunov equation~\cite{DPRL2007}%
\begin{equation}
\mathbf{AV}+\mathbf{VA}^{T}=-\mathbf{Q,}
\end{equation}%
where we introduce the noise correlation matrix defined by the elements $\mathbf{Q}_{n,m}=\frac{1}{2}\langle
\mathbf{\hat{N}}_{n}(t)\mathbf{\hat{N}}_{m}(t^{\prime })+\mathbf{\hat{N}}_{m}(t)\mathbf{\hat{N}}_{n}(t^{\prime })\rangle $. For our considered case, the matrix $\mathbf{Q}$ can be obtained as
\begin{eqnarray}
\mathbf{Q} & = & \text{diag}[\kappa ,\kappa ,\gamma _{x_{1}}( 2\bar{n}_{x_{1}}^{th}+1),\gamma _{x_{2}}( 2\bar{n}_{x_{2}}^{th}+1) ,
\notag \\
&&\gamma _{y_{1}}( 2\bar{n}_{y_{1}}^{th}+1),\gamma _{y_{2}}(2\bar{n}_{y_{2}}^{th}+1) ,\gamma _{z_{1}}( 2\bar{n}_{z_{1}}^{th}+1), \notag \\
&&\gamma _{z_{2}}(2\bar{n}_{z_{2}}^{th}+1) , 0,0 ,0,0,0 ,0 ].
\end{eqnarray}%

\begin{figure}[t!]
\centering\includegraphics[scale=1]{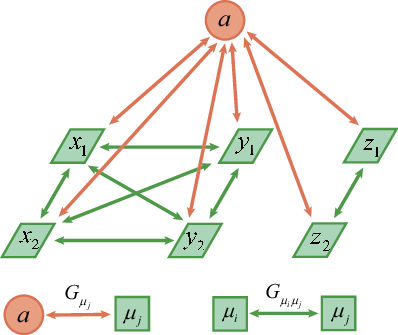}
\caption{{Schematic of the seven-mode coupling configuration for the coupled cavity-levitated NP system, including the optomechanical couplings $G_{\mu_{j}}$ ($\mu=x,y,z$ and $i,j=1,2$) between the cavity mode (orange circle) and the motional modes of the two NPs (green squares) along the three spatial
directions, as well as the mechanical couplings $G_{\mu_{i}\mu_{j}}$ between the motional modes of the two NPs in three directions. Notably, the motional mode along the $Z$-direction couples only to the mode along the same
direction.}}
\label{modelv2}
\end{figure}%

Based on the linearized Langevin equations~(\ref{30}), we can see that the system is reduced to a linear seven bosonic-mode network consisting of the cavity mode $a$ and six mechanical modes $\mu_{j}$ for $\mu=x,y,z$ and $j=1,2$. Figure~\ref{modelv2} shows the coupling configuration of the linearized system. Here, we can see that the optomechanical couplings exist between the cavity mode and the motional modes of the two NPs along three directions. Moreover, the mechanical couplings exist between the motional modes of the two NPs in three directions. While all mechanical modes are coupled to the cavity mode, the $Z$-direction motional modes of the two NPs are only coupled with each other, in contrast to the cross-coupling among these $X$- and $Y$-direction modes.

\begin{figure}[t!]
\centering\includegraphics[width=0.48\textwidth]{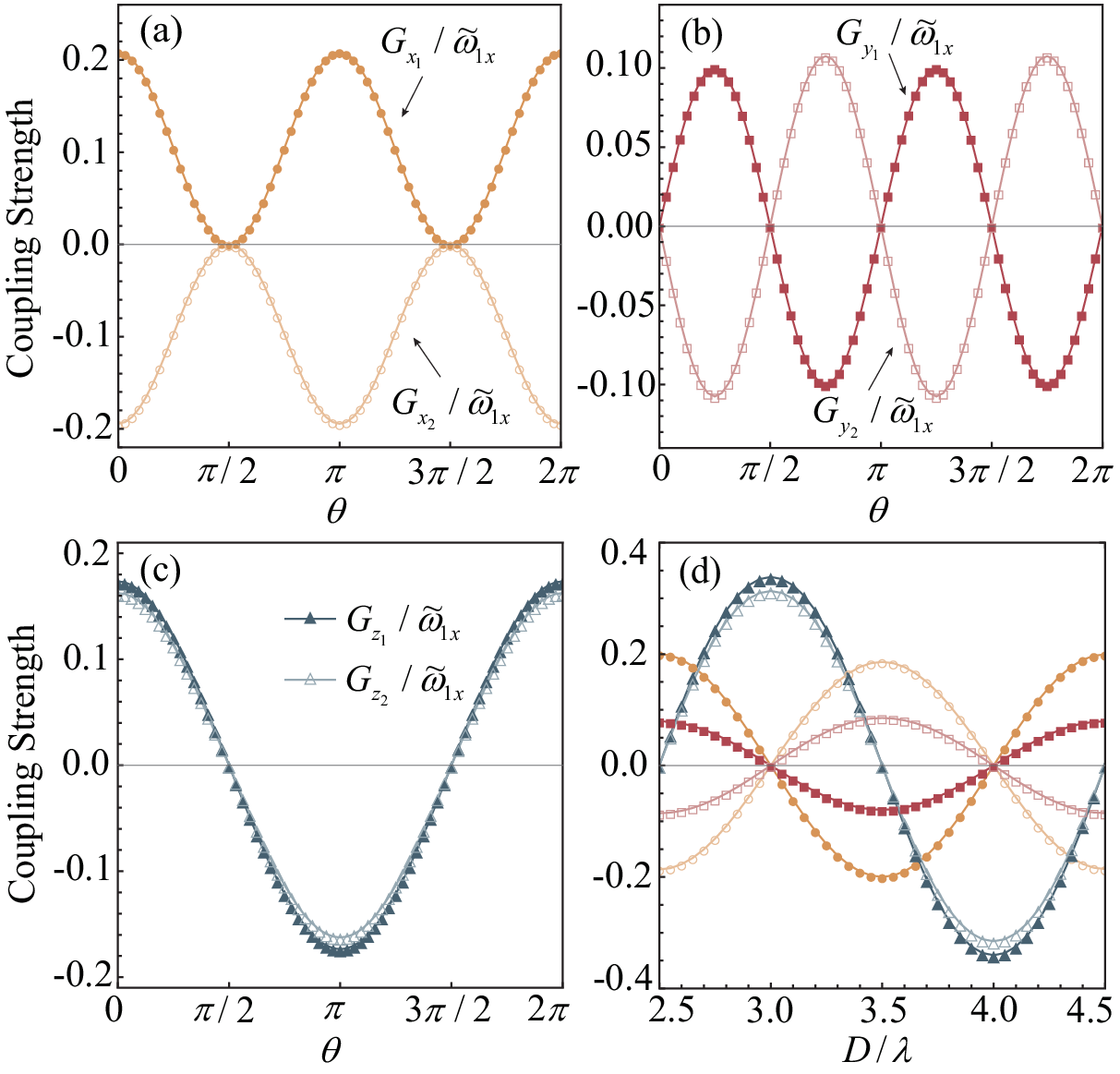}
\caption{{The scaled coupling strengths (a) $G_{x_{j}}/\tilde{\omega}_{1x}$, (b) $G_{y_{j}}/\tilde{\omega}_{1x}$, and (c) $G_{z_{j}}/\tilde{\omega}_{1x}$ ($j=1,2$) as functions of the polarization angle $\theta$ at the distance $D= 2.65\lambda $ between the two NPs. (d) The scaled coupling strengths $G_{\mu_{j}}/\tilde{\omega}_{1x}$ ($\mu=x,y,z$) as functions of the NP spacing $D/\lambda$ at the polarization angle $\theta=\pi/8$. Other parameters used include: the radius of silica NP is $r_{0}=70$ nm, $\lambda =1064$ nm, $P_{\text{tw}_{1}}=0.8$ W, $P_{\text{tw}_{2}}=0.6$ W, $\tilde{\Delta} /\tilde{\omega}_{1x}=1$, and $\kappa /\tilde{\omega}_{1x}=0.2.$}}
\label{modelv3}
\end{figure}

To throughly investigate the coupling effect, we focus on two crucial parameters, namely the polarization angle $\theta $ and the distance $D$ between the two NPs. Accordingly, we present the coupling strengths $G_{\mu_{j}}/\tilde{\omega}_{1x}$ as functions of $\theta $ and $D$, as shown in Fig.~\ref{modelv3}. Here, we find that the scaled coupling strengths $G_{x_{j}}/\tilde{\omega}_{1x}$ and $G_{y_{j}}/\tilde{\omega}_{1x}$ change periodically as the polarization angle $\theta $ with a period $\pi $. Particularly, when $\theta = \pi/2$ or $3\pi/2$, the cavity field and the optical tweezer field are decoupled from each other. At these angles, the coupling strengths in three directions are zero. Whereas when $\theta = 0$ or $\pi$, the motional modes of the two particles in the $Y$-direction are decoupled from the cavity mode $( G_{y_{j}}/\tilde{\omega}_{1x}\approx 0) $. This confirms that the polarization angle $\theta $ can be used to control the coupling channels in specific dimension, which is consistent with the theoretical expectation. By tuning the polarization angle $\theta $ away from these values, the $Y$-axis coupling is activated,
thereby achieving fully-connected coupling in the 3D direction. Additionally, we investigate the coupling strengths $G_{\mu_{j}}/\tilde{\omega}_{1x}$ for $\mu=x,y,z$ and $j=1,2$ as functions of the particle spacing $D$ in Fig.~\ref{modelv3}(d). It can be found that the coupling strengths in the three directions vary periodically with $D$. For the $X$- and $Y$-directions, the coupling strengths $G_{x_{j}}/\tilde{\omega}_{1x}$ and $G_{y_{j}}/\tilde{\omega}_{1x}$ reach a maximum at $D\approx 2.5\lambda $, while at this distance the coupling strength $G_{z_{j}}/\tilde{\omega}_{1x}$ in the $Z$-direction is zero. When $D\approx 3\lambda $, the coupling strengths in the $X$- and $Y$-directions become zero, whereas that in the $Z$-direction reaches its maximum. In our following calculations, we select $D= 2.65\lambda $ for considerably strong couplings.

\section{SIMULTANEOUS GROUND-STATE COOLING OF SIX MECHANICAL MODES}\label{SIMULTANEOUS GROUND-STATE COOLING OF SIX MECHANICAL MODES}
To evaluate the simultaneous 3D ground-state cooling, we study the cooling performance of these six mechanical modes by calculating the final mean phonon numbers%
\begin{equation}
\bar{n}_{\mu_{j}}=\frac{1}{2}[ \langle \delta
\hat{q}_{\mu_{j}}^{2}\rangle +\langle \delta \hat{p}_{\mu_{j}}^{2}\rangle -1%
] ,
\end{equation}%
where $\mu=x,y,z$ and $j=1,2$. These stationary variances of the mechanical modes are given by the
corresponding diagonal matrix elements of the covariance matrix,%
\begin{subequations}
\begin{align}
\bar{n}_{x_{1}}&=\frac{1}{2}(\mathbf{V}_{9,9} +\mathbf{V}_{3,3}-1), \\
\bar{n}_{x_{2}}&=\frac{1}{2}(\mathbf{V}_{10,10} +\mathbf{V}_{4,4}-1), \\
\bar{n}_{y_{1}}&=\frac{1}{2}(\mathbf{V}_{11,11} +\mathbf{V}_{5,5}-1), \\
\bar{n}_{y_{2}}&=\frac{1}{2}(\mathbf{V}_{12,12} +\mathbf{V}_{6,6}-1), \\
\bar{n}_{z_{1}}&=\frac{1}{2}(\mathbf{V}_{13,13} +\mathbf{V}_{7,7}-1), \\
\bar{n}_{z_{2}}&=\frac{1}{2}(\mathbf{V}_{14,14} +\mathbf{V}_{8,8}-1).
\end{align}
\end{subequations}

\begin{figure}[t!]
\center\includegraphics[width=0.48\textwidth]{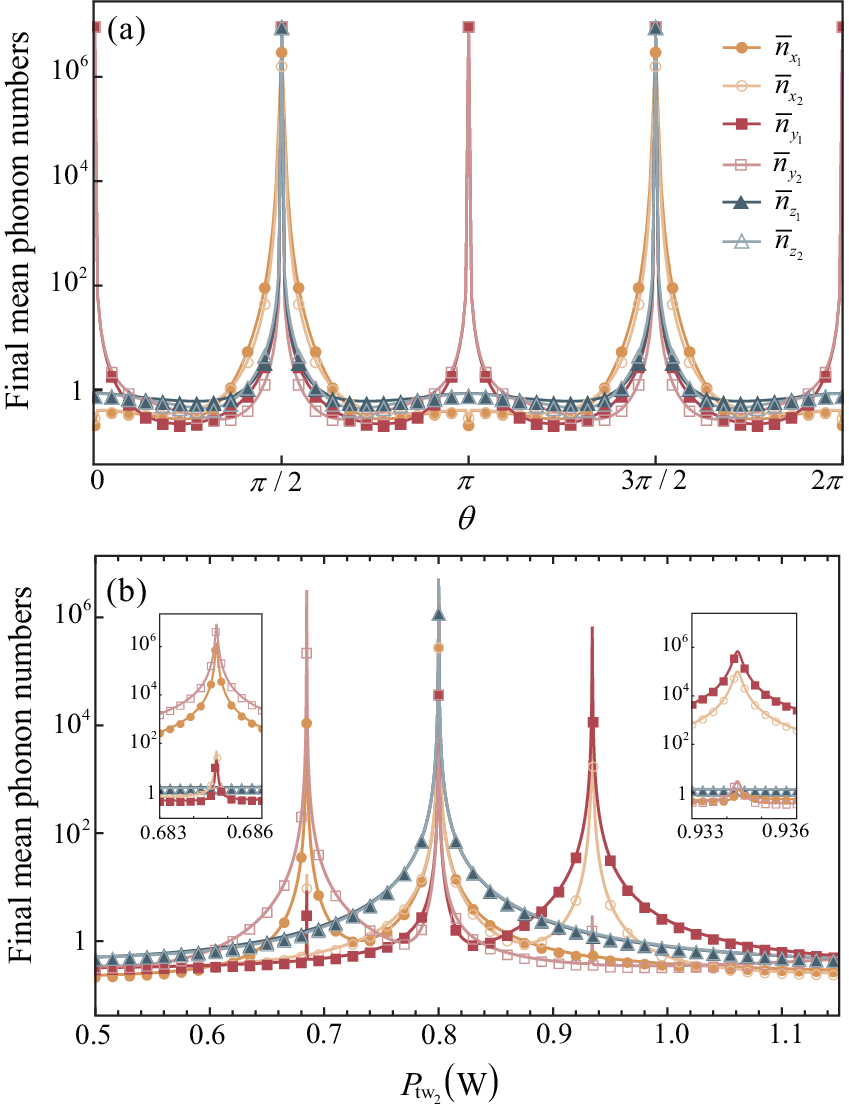}
\caption{{(a) The final mean phonon numbers $\bar{n}_{\mu_{j}}$ ($\mu=x,y,z$ and $j=1,2$) versus the polarization angle $\theta$ at $P_{\text{tw}_{2}}=0.6$~W. (b) The final mean phonon numbers $\bar{n}_{\mu_{j}}$ versus the power $P_{\text{tw}_{2}}$ of the tweezer 2 at $\theta=\pi/8 $. The two insets are zoom-in plots of the phonon numbers in a narrower range. Other parameters used are as follows: the silica nanoparticle of radius is $r_{0}= 70$ nm, the separation of the particles is $D = 2.65\lambda $, $\lambda  =1064$~nm, $P_{\text{tw}_{1}}=0.8$~W, $\tilde{\Delta} /\tilde{\omega}_{1x}=0.7,$ $\kappa /\tilde{\omega}%
_{1x}=0.2,$ $\bar{n}_{\mu_{j}}^{th}=10^{7},$ and $\gamma_{\mu_{j}} /\tilde{\omega}_{1x} =0.5\times 10^{-9}$.}}
\label{modelv4}
\end{figure}

As shown in Figs.~\ref{modelv3}, these coupling strengths depend on the polarization angle $\theta$. Therefore, we investigate the dependence of the cooling performance on the angle $\theta$. Figure~\ref{modelv4}(a) shows the final mean phonon numbers $\bar{n}_{\mu_{j}}$ for $\mu=x,y,z$ and $j=1,2$ as functions of the polarization angle $\theta$. It is found that the cooling of these six modes is periodic in $\theta$ with a period of $\pi$. When $\theta = \pi/2$, the two fields are orthogonal, leading to the decoupling of both the optical and mechanical modes. When $\theta = 0$ or $\pi$, the $y_{1}$ and $y_{2}$ modes of the two nanoparticles are completely decoupled from the cavity mode ($G_{y_{j}}/\tilde{\omega}_{1x}=0$). As a result, the cooling channel for the $y$ modes is closed, and the $y$ modes cannot be efficiently cooled. In contrast, when $\theta \approx \pi/4$, the simultaneous cooling of these six modes achieves optimal performance. Note that at $\theta = \pi$, the coupling strengths $G_{x_{1}}/\tilde{\omega}_{1x}$ and $G_{x_{2}}/\tilde{\omega}_{1x}$ reach their maxima. Therefore, the cooling effect of the $x_{1}$ and $x_{2}$ modes is the best at this point.

To investigate the effect of the optical tweezers power on the cooling of these mechanical modes, we plot the final mean phonon numbers $\bar{n}_{\mu_{j}}$ for $\mu=x,y,z$ and $j=1,2$ as functions of the power $P_{\text{tw}_{2}}$ of the tweezer 2 when $\theta =\pi /8$ and $P_{\text{tw}_{1}}=0.8$~W, as shown in Fig.~\ref{modelv4}(b). We find that, when the two optical tweezers have the same power $P_{\text{tw}_{1}}=P_{\text{tw}_{2}}$%
, the cooling of these six mechanical modes is strongly suppressed. When $%
P_{\text{tw}_{2}}\approx 0.6847$ W or $P_{\text{tw}_{2}}\approx 0.9343$ W, the final phonon numbers of the $X$- and $Y$-directional modes cannot be effectively decreased, while the $z$ mode remains unaffected. We can explain these phenomena by analyzing the dark-mode effect in this system~\cite{CNJP2008,CVS2012,YAPRL2012,LAPRL2012,DJXPRA2020,Lai2020,JAFPRL2022,JDCPRA2022,Huang2023,Jarx2023}.

Based on the linearized Langevin equations [Eq.~(\ref{30})], we can derive an effective Hamiltonian as follows,%
\begin{eqnarray}
\hat{H}_{\text{lin}}/\hbar  &=&\Delta ^{\prime }\delta \hat{a}^{\dag }\delta
\hat{a}+\sum_{j=1,2}\sum_{\mu=x,y,z}\frac{\tilde{\omega}_{j\mu}}{2}( \delta
\hat{p}_{\mu_{j}}^{2}+\delta \hat{q}_{\mu_{j}}^{2})   \nonumber \\
&&+\sum_{j=1,2}( \tilde{G}_{ax_{j}}\delta \hat{a}+\tilde{G}_{ax_{j}}^{\ast
}\delta \hat{a}^{\dag }) \delta \hat{q}_{x_{j}}   \nonumber \\
&&+\sum_{j=1,2}( \tilde{G}_{ay_{j}}\delta \hat{a}+\tilde{G}_{ay_{j}}^{\ast }\delta \hat{a}^{\dag })
\delta \hat{q}_{y_{j}}  \nonumber \\
&&+\sum_{j=1,2}i( \tilde{G}_{az_{j}}\delta \hat{a}-\tilde{G}_{az_{j}}^{\ast }\delta
\hat{a}^{\dag }) \delta \hat{q}_{z_{j}}\nonumber \\
&&-G_{x}\delta \hat{q}_{x_{1}}\delta \hat{q}_{x_{2}}
-G_{y}\delta \hat{q}_{y_{1}}\delta \hat{q}_{y_{2}}-G_{z}\delta \hat{q}_{z_{1}}\delta
\hat{q}_{z_{2}}  \nonumber \\
&&+G_{x_{1}y_{1}}\delta \hat{q}_{x_{1}}\delta \hat{q}_{y_{1}}+G_{x_{2}y_{2}}\delta \hat{q}_{x_{2}}\delta \hat{q}_{y_{2}}  \nonumber \\
&&-G_{x_{1}y_{2}}\delta
\hat{q}_{x_{1}}\delta \hat{q}_{y_{2}}-G_{x_{2}y_{1}}\delta \hat{q}_{x_{2}}\delta \hat{q}_{y_{1}},
\end{eqnarray}%
where the linearized optomechanical coupling strengths are given by $\tilde{G}_{a\mu_{j}}=%
G_{\mu_{j}}^{\prime }+G_{a\mu_{j}}^{\prime }.$ To analyze the dark-mode effect, we introduce the creation and annihilation operators $\hat{b}_{\mu_{j}}^{\dag }=( \hat{q}_{\mu_{j}}-i\hat{p}_{\mu_{j}}) /\sqrt{2%
}$ and $\hat{b}_{\mu_{j}}=(  \hat{q}_{\mu_{j}}+i \hat{p}_{\mu_{j}}) /\sqrt{2}$ for these mechanical modes. By transforming the Hamiltonian into the $\{\hat{a},\hat{b}_{\mu_{j}}\}$ representation and ignoring the counter-rotating terms, we obtain the new Hamiltonian under the rotating-wave approximation as
\begin{eqnarray}
\hat{H}'/\hbar  &\approx &\Delta ^{\prime }\hat{a}^{\dag }
\hat{a}+\sum_{j=1,2}\sum_{\mu=x,y,z} \tilde{\omega}_{j\mu}  \hat{b}_{\mu_{j}}^{\dag } \hat{b}_{\mu_{j}}   \nonumber \\
&&+\sum_{j=1,2} \frac{1}{\sqrt{2}}( \tilde{G}_{ax_{j}} \hat{a} \hat{b}_{x_{j}}^{\dag } +\tilde{G}_{ax_{j}}^{\ast
} \hat{a}^{\dag } \hat{b}_{x_{j}} )  \nonumber \\
&&+\sum_{j=1,2} \frac{1}{\sqrt{2}}( \tilde{G}_{ay_{j}} \hat{a} \hat{b}_{y_{j}}^{\dag } +\tilde{G}_{ay_{j}}^{\ast
} \hat{a}^{\dag } \hat{b}_{y_{j}} )  \nonumber \\
&&+\sum_{j=1,2} \frac{i}{\sqrt{2}} ( \tilde{G}_{az_{j}} \hat{a} \hat{b}_{z_{j}}^{\dag } -\tilde{G}_{az_{j}}^{\ast
} \hat{a}^{\dag } \hat{b}_{z_{j}} ) \nonumber \\
&&-\sum_{\mu=x,y,z} \frac{1}{2} G_{\mu}(  \hat{b}_{\mu_{1}}^{\dag } \hat{b}_{\mu_{2}} +\hat{b}_{\mu_{1}} \hat{b}_{\mu_{2}}^{\dag } )
\nonumber \\
&&+ \sum_{j=1,2}\frac{1}{2} G_{x_{j}y_{j}}(  \hat{b}_{x_{j}}^{\dag } \hat{b}_{y_{j}} +\hat{b}_{x_{j}} \hat{b}_{y_{j}}^{\dag } )  \nonumber \\
&&- \sum_{j=1,2}\frac{1}{2} G_{x_{j}y_{\bar{j}}}(  \hat{b}_{x_{j}}^{\dag } \hat{b}_{y_{\bar{j}}} +\hat{b}_{x_{j}} \hat{b}_{y_{\bar{j}}}^{\dag } ) . \label{H'}
\end{eqnarray}%
For analysis convenience, the Hamiltonian in Eq.~(\ref{H'}) of the optomechanical network can be rewritten as~\cite{Jarx2023}
\begin{equation}
\hat{H}'/\hbar =( \hat{a}^{\dag },\mathbf{\hat{b}}^{\dag }) \mathbf{H}_{ab}(
\hat{a},\mathbf{\hat{b}}) ^{\text{T}},
\end{equation}%
where $\mathbf{\hat{b}=}( \hat{b}_{x_{1}},\hat{b}_{x_{2}},\hat{b}_{y_{1}},\hat{b}_{y_{2}},\hat{b}_{z_{1}},\hat{b}_{z_{2}})
^{\text{T}},$
\begin{figure*}[t!]
\centering\includegraphics[width=0.96\textwidth]{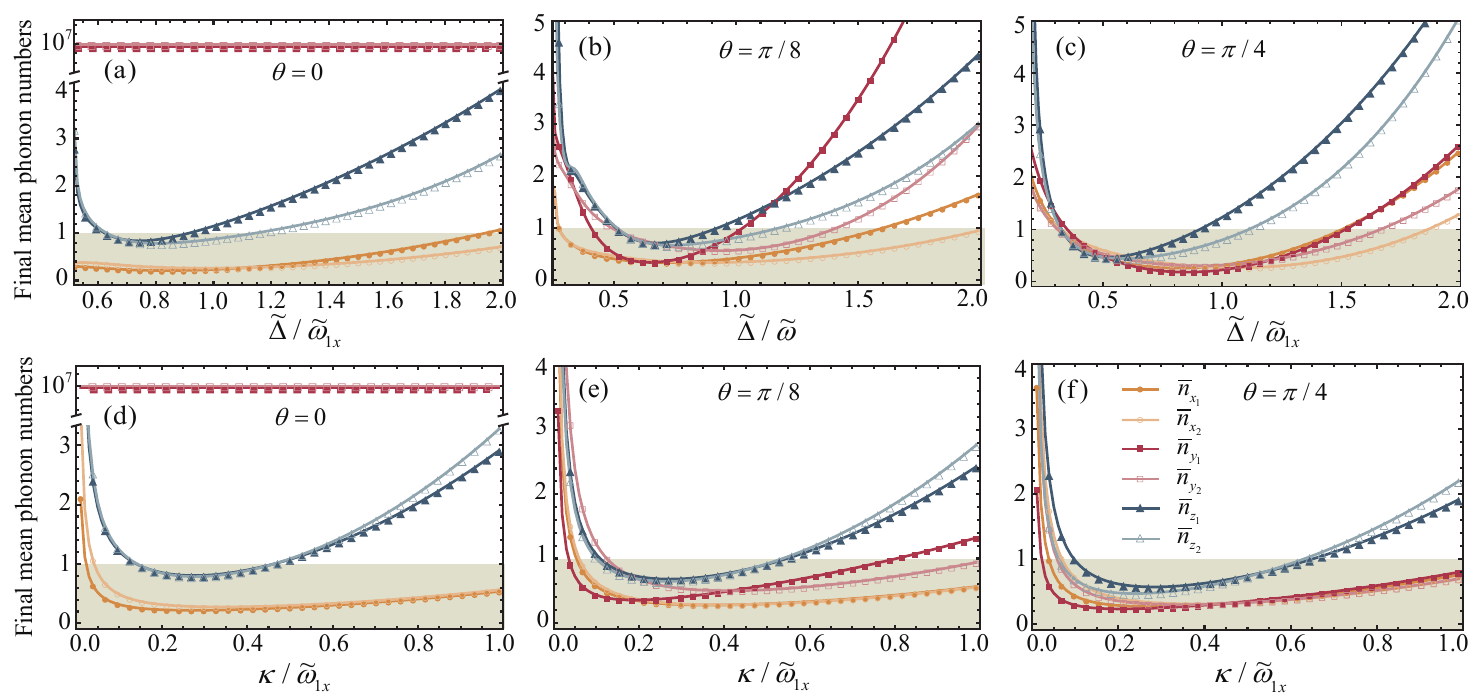}
\caption{{The final mean phonon numbers $\bar{n}_{\mu_{j}}$ ($\mu=x,y,z$ and $j=1,2$) for these six mechanical modes as functions of the scaled driving detuning $\tilde{\Delta} /\tilde{\omega}_{1x}$ for $\kappa /\tilde{\omega}_{1x}=0.2 $ at different polarization angles: (a) $\theta =0,$ (b) $\theta =\pi /8$, and (c) $\theta =\pi /4$. The final mean phonon numbers $\bar{n}_{\mu_{j}}$ for these six mechanical modes as functions of the scaled cavity linewidth $\kappa /\tilde{\omega}_{1x}$ for $\tilde{\Delta}/\tilde{\omega}_{1x}=0.7 $ at different polarization angles: (d) $\theta =0,$ (e) $\theta =\pi /8$, and (f) $\theta =\pi /4$. Other parameters used are as follows: the silica nanoparticle of radius is $r_{0}=70$ nm, the separation of the particles is $D=2.65\lambda $, $\lambda =1064$ nm, $P_{\text{tw}_{1}}=0.8$ W, $P_{\text{tw}_{2}}=0.6$ W, $\bar{n}_{\mu_{j}}^{th}=10^{7}$, and $\gamma _{\mu_{j}}/\tilde{\omega}_{1x}=0.5\times 10^{-9}$.}}
\label{modelv5}
\end{figure*}%
and the coefficient matrix is given by
\begin{equation} \label{Hab}
\mathbf{H}_{ab}=%
\begin{pmatrix}
\mathbf{H}_{a} & \mathbf{C}_{ab} \\
\mathbf{C}_{ab}^{\dag } & \mathbf{H}_{b}%
\end{pmatrix} ,
\end{equation}%
with $\mathbf{H}_{a}=\tilde{\Delta}$, $\mathbf{C}_{ab}/\sqrt{2}=( \tilde{G}_{ax_{1}},%
\tilde{G}_{ax_{2}},\tilde{G}_{ay_{1}},\tilde{G}_{ay_{2}}, i \tilde{G}_{az_{1}}, i \tilde{G}_{az_{2}}) $, and%
\begin{equation} \label{Hb}
\mathbf{H}_{b}=%
\begin{pmatrix}
\tilde{\omega}_{1x} & -\frac{G_{x}}{2} & \frac{G_{x_{1}y_{1}}}{2} & -\frac{%
G_{x_{1}y_{2}}}{2} & 0 & 0 \\
-\frac{G_{x}}{2} & \tilde{\omega}_{2x} & -\frac{G_{x_{2}y_{1}}}{2} & \frac{%
G_{x_{2}y_{2}}}{2} & 0 & 0 \\
\frac{G_{x_{1}y_{1}}}{2} & -\frac{G_{x_{2}y_{1}}}{2} & \tilde{\omega}_{1y} &
-\frac{G_{y}}{2} & 0 & 0 \\
-\frac{G_{x_{1}y_{2}}}{2} & \frac{G_{x_{2}y_{2}}}{2} & -\frac{G_{y}}{2} &
\tilde{\omega}_{2y} & 0 & 0 \\
0 & 0 & 0 & 0 & \tilde{\omega}_{1z} & -\frac{G_{z}}{2} \\
0 & 0 & 0 & 0 & -\frac{G_{z}}{2} & \tilde{\omega}_{2z}%
\end{pmatrix}%
.
\end{equation}
We first diagonalize the mechanical-mode subnetwork [Eq.~(\ref{Hb})]. By introducing a unitary transformation $\mathbf{U}_b$, we have $\mathbf{U}_b \mathbf{H}_b \mathbf{U}_b^\dagger = \mathbf{H}_B = \mathrm{diag}[\tilde{\omega}_{1}, \tilde{\omega}_{2}, \dots, \tilde{\omega}_{6}]$, so that the original mechanical modes $\mathbf{\hat{b}}$ are converted into normal modes $\mathbf{\hat{B}}=(\hat{B}_1,\hat{B}_2,\dots,\hat{B}_6) = \mathbf{U}_{b} \mathbf{\hat{b}}$. The cavity-mode subnetwork $\mathbf{H}_{a}$ is diagonal (as it consists of a single cavity mode), so we have $\mathbf{H}_{A}=\mathbf{H}_{a}$. In the normal-mode representation, the system is reduced to a bipartite-graph network, with the coupling matrix $\mathbf{C}_{AB} = \mathbf{C}_{ab} \mathbf{U}_b^\dagger =( \tilde{G}_{aB_{1}},\tilde{G}_{aB_{2}},\dots,\tilde{G}_{aB_{6}})$. In the normal mode representation, the coefficient matrix $\mathbf{H}_{AB}$ of the system can be expressed as an arrowhead matrix
\begin{equation}
\mathbf{H}_{AB}=%
\begin{pmatrix}
\mathbf{H}_{A} & \mathbf{C}_{AB} \\
\mathbf{C}_{AB}^{\dag } & \mathbf{H}_{B}%
\end{pmatrix} .
\end{equation}%

For the case of $P_{\text{tw}_{1}}=P_{\text{tw}_{2}}$, it can be found that the frequencies of the mechanical modes in the same direction for the two NPs are identical: $\tilde{\omega}_{1x}=\tilde{\omega}%
_{2x}$, $\tilde{\omega}_{1y}=\tilde{\omega}_{2y}$, and $\tilde{\omega}_{1z}=%
\tilde{\omega}_{2z}$. Meanwhile, the coupling strengths satisfy the
relationships $\tilde{G}_{ax_{1}}=-\tilde{G}_{ax_{2}},$ $\tilde{G}_{ay_{1}}=-\tilde{G}_{ay_{2}},
$ $\tilde{G}_{az_{1}}=-\tilde{G}_{az_{2}},$ and $%
G_{x_{1}y_{1}}=G_{x_{2}y_{2}}=G_{x_{1}y_{2}}=G_{x_{2}y_{1}}.$ According to the dark-mode theorem~\cite{Jarx2023}, if the $s$th ($s=1$-$6$) element of the coupling matrix $\mathbf{C}_{AB}$ is zero, the corresponding normal mode $\hat{B}_{s}$ becomes a dark mode. Through numerical calculations, we find that when $P_{\text{tw}_{1}}=P_{\text{tw}_{2}}=0.8$~W, the coupling matrix elements $\tilde{G}_{aB_{1}}=\tilde{G}_{aB_{3}}=\tilde{G}_{aB_{6}}=0$ in $\mathbf{C}_{AB}$, which means that there exist three dark modes $\hat{B}_{1}$, $\hat{B}_{3}$, and $\hat{B}_{6}$. Therefore, the simultaneous ground-state cooling of these six mechanical modes will be significantly suppressed.

For the case of $P_{\text{tw}_{2}} \approx 0.6847$~W, we find that $\tilde{\omega}_{1x}=\tilde{\omega}_{2y}.$ According to the dark-mode theorem~\cite{Jarx2023}, if all coupling elements are nonzero but $l$ mechanical normal modes are degenerate in the arrowhead matrix, then within this degenerate subspace there exists one bright mode and $l-1$ dark modes. Through numerical calculations, we find that none of these coupling elements are zero in the arrowhead matrix, but in $\mathbf{H}_B$, $\tilde{\omega}_{2}/\tilde{\omega}_{1x} \approx\tilde{\omega}_{3}/\tilde{\omega}_{1x} \approx 1.0$, i.e., $l=2$, so there is one dark mode in the system. Similarly, when $P_{\text{tw}_{2}} \approx 0.9343$~W, we have $\tilde{\omega}_{2}/\tilde{\omega}_{1x}\approx \tilde{\omega}_{3}/\tilde{\omega}_{1x} \approx 1.1$, then there is one dark mode. Note that in the arrowhead matrix $\mathbf{H}_{AB}$, only the matrix elements associated with the $x_{j}$ and $y_{j}$ modes are degenerate. As a result, dark modes exist only in the $x_{j}$, $y_{j}$-mode subspace. However, the $z_{j}$ modes are nondegenerate with other mechanical modes, so there are no dark modes involving them, and their cooling process remains unaffected.

As the cavity detuning and linewidth play crucial roles in the simultaneous cooling of these mechanical modes, we plot in Fig.~\ref{modelv5} the final mean phonon numbers $\bar{n}_{x_{1}}$, $\bar{n}_{x_{2}},$ $\bar{n}_{y_{1}},$ $\bar{n}_{y_{2}},$ $\bar{n}_{z_{1}},$ and $\bar{n}_{z_{2}}$ of these six mechanical modes (corresponding to the $X$-, $Y$-, and $Z$-direction motions of the two NPs) as functions of the scaled driving detuning $\tilde{\Delta} /\tilde{\omega}_{1x}$ and the scaled cavity linewidth $\kappa /\tilde{\omega}_{1x}$. As shown in Figs.~\ref{modelv3}, when $\theta = 0$ or $\pi$, the $y_{1}$ and $y_{2}$ modes of the two NPs are completely decoupled from the cavity mode ($G_{y_{j}}/\tilde{\omega}_{1x}=0$). As a result, the cooling channel for the $y_{j}$ modes is closed and the $y_{j}$ modes cannot be efficiently cooled. As shown in Figs.~\ref{modelv5}(a) and ~\ref{modelv5}(d), the ground-state cooling of $x_{j}$ and $z_{j}$ modes can be achieved in this case, while the phonon numbers of the $Y$-directional modes remain significantly higher than those of other modes, thereby failing to achieve the ground-state cooling. When $\theta \neq 0$ (e.g., $\pi /8$ and $\pi /4$), the coupling channel along the $Y$-direction is activated, enabling simultaneous 3D ground-state cooling of the system, as shown in Figs.~\ref{modelv5}(b), ~\ref{modelv5}(c), ~\ref{modelv5}(e), and ~\ref{modelv5}(f). Furthermore, as $\theta $ increases from $\pi /8$ to $\pi /4$, the final mean phonon numbers of these six mechanical modes exhibit a decreasing trend, indicating a gradual improvement in the overall cooling performance. At $\theta=\pi /4$, the phonon numbers in three directions reach their minimum.

In Figs.~\ref{modelv5}(a-c), we observe that for the scaled cavity linewidth $\kappa /\tilde{\omega}_{1x}=0.2$, the final phonon number of each mechanical mode varies with the scaled driving detuning $\tilde{\Delta} /\tilde{\omega}_{1x}$. In general, when the driving detuning resonates with the mechanical mode (i.e., the red sideband resonance $\tilde{\Delta}\approx \tilde{\omega}_{j\mu}$), the optimal cooling effect can be achieved. However, due to the couplings among different mechanical modes and between these modes and the optical field, the effective frequencies of the modes experience small shifts, causing the optimal cooling point to deviate from the bare resonance frequency. In addition, the polarization angle $\theta $ also affects the eigenfrequencies of the mechanical modes, leading to different optimal working points for different $\theta $. Considering the cooling performance of all modes comprehensively, we choose $\tilde{\Delta} /\tilde{\omega}_{1x}=0.7$ as a unified working point for the subsequent analysis of the scaled cavity linewidth dependence, where the six modes achieve reasonably good cooling.

Figures~\ref{modelv5}(d-f) show the final phonon numbers as functions of the scaled cavity linewidth $\kappa /\tilde{\omega}_{1x}$ for the scaled driving detuning $\tilde{\Delta} /\tilde{\omega}_{1x}=0.7$. Overall, the cooling performance is optimal in the resolved-sideband regime ($\kappa /\tilde{\omega}_{1x} < 1$), which is a feature of the resolved-sideband cooling. Consistent with Figs.~\ref{modelv5}(a-c), as $\theta $ increases from $0$ to $\pi /4,$ the final phonon numbers of all modes in the optimal parameter region show a decreasing trend, further confirming that increasing the polarization angle $\theta $ enhances the simultaneous 3D cooling effect. Notably, at $\theta =\pi /4,$ $\tilde{\Delta} /\tilde{\omega}_{1x}=0.7$, and $\kappa /\tilde{\omega}_{1x}=0.2$, the phonon numbers of these six mechanical modes are reduced to below $1$, demonstrating that simultaneous ground-state cooling in three spatial dimensions of the particles can be achieved via the polarization-angle-controlled coherent scattering mechanism.

\section{Discussions and Conclusion}\label{conclusion and discussions}
Finally, we present some discussions on the experimental implementation of the present scheme. In this work, we considered the coupled cavity-levitated-nanoparticle system in which two levitated nanoparticles are trapped within a cavity, and the two nanoparticles can be controlled independently. This physical configuration can be realized by current experiments, because two nanoparticles trapped by separate optical tweezers within a single cavity have been realized using current experimental techniques~\cite{JVarx2023,AQAPRL2025}. In particular, the polarizations of the two optical tweezers can be independently controlled, enabling flexible tuning of the trapping potentials and dipole orientations~\cite{Varx2023,Marx2023,JMS2022,YEKO2018}.

In our simulations, we considered the silica NPs ($r_0=70$~nm) with a real polarizability $\alpha$ and a negligible absorption effect~\cite{DPRL2019}. The center-of-mass mechanical frequencies of the trapped nanoparticles are $\tilde{\omega}_{1x,1y,1z}/2\pi \approx (406,439,154)$~kHz and $\tilde{\omega}_{2x,2y,2z}/2\pi \approx (351,380,133)$~kHz at $P_{\text{tw}_{1}}=0.8$~W and $P_{\text{tw}_{2}}=0.6$~W, the cavity length $L = 1.07$~cm and waist $w_{c}=41.1$~$\mu$m, all these parameters are consistent with typical parameters reported in levitated optomechanics experiments~\cite{UPRL2019}. For computational simplicity, we assumed that all these mechanical modes of both nanoparticles have the same damping rate $\gamma_{\mu_j}/\tilde{\omega}_{1x}\approx 10^{-9} $. Such a damping rate is achievable in levitated particle systems. In the ultra-high vacuum, the center-of-mass motion of the dielectric sphere is well decoupled from the environment, and the $Q$ factors can approach $10^{12}$~\cite{JTR2020,Gaxv2307}. Note that the mechanical quality factor $Q$ in the levitated particle systems does not increase indefinitely as the gas pressure decreases. The effective damping of the system originates from collisions with background gas molecules and photon recoil heating. When the gas pressure decreases, the gas-induced damping decreases, while photon recoil heating gradually becomes the dominant source of decoherence, thus setting the limit on the maximum achievable quality factor~\cite{VJCPRL2016}. In this paper, we focus our discussions on the regime where the decoherence is dominated by gas collisions, in which the system dynamics can be described by the Langevin equations. Based on the above discussions, the proposed model should be experimentally accessible with the state-of-the-art technology.

In conclusion, we have theoretically investigated the simultaneous cooling of six mechanical modes of two nanoparticles in a coupled cavity-levitated-nanoparticle system via coherent scattering. We have derived the Hamiltonian of the system and demonstrated that the system can be simplified to a seven-mode linearized optomechanical network. We have found that the coupling strengths between the cavity mode and six mechanical modes are highly sensitive to the polarization angle $\theta$ between the cavity field and the tweezer fields. Specifically, when $\theta = \pi/2$ or $\theta=3\pi/2$, the two fields are perpendicular and lead to the decoupling of both optical and mechanical modes. Meanwhile, when $\theta = 0$ or $\pi$, the $Y$-direction motional modes decouple from both the cavity modes and other motional modes of the nanoparticles. We have also studied the ground-state cooling of the mechanical modes. Our results reveal that the presence of dark modes will suppress the simultaneous ground-state cooling of these mechanical modes. Beyond the symmetric case of equal tweezer powers, we further utilize the arrowhead-matrix method to uncover the dark-mode effect in the system. Furthermore, when the dark-mode effect is broken, we demonstrate that both the $X$- and $Z$-direction motions can be significantly cooled at $\theta=0$, whereas the simultaneous 3D ground-state cooling of the two nanoparticles can be realized around $\theta=\pi/8$ and $\pi/4$. This work paves the way for exploring collective macroscopic quantum effects in levitated multiparticle systems, and suggests new strategies for the implementation of tunable optomechanical platforms.

\begin{acknowledgments}
J.-Q.L. was supported in part by National Natural Science Foundation of China (Grants No.~12575015, No.~12247105, and No.~12421005), National Key Research and Development Program of China (Grant No.~2024YFE0102400), and Hunan Provincial Major Sci-Tech Program (Grant No.~2023ZJ1010).
\end{acknowledgments}

\appendix*
\begin{widetext}
\section{Derivation of the Hamiltonians in Eq.~(\ref{20})} \label{Appendix}
In this Appendix, we present the detailed derivation of the Hamiltonians in Eq.~(\ref{20}). In Eq.~(\ref{20}), there are the kinetic energy term $\sum_{j=1,2}\mathbf{\hat{P}}_{j}^{2}/2m_{j}$, and the free Hamiltonian $\hbar \Delta ^{\prime }\hat{a}^{\dag }\hat{a}$ of the cavity field in the rotating frame, as well as the interaction Hamiltonian terms. The interaction Hamiltonian part can be obtained by expanding the square terms in Eq.~(\ref{7}). As we explained in the paragraph below Eq.~(\ref{7}), the high-order scattering terms should be discarded in Eq.~(\ref{7}). Therefore, the interaction Hamiltonian in Eq.~(\ref{7}) is approximately reduced to~\cite{YYPRA2024}%
\begin{align}
\hat{H}_{\text{int}} &\approx \sum_{j=1,2} \left[ \hat{H}_{\text{tw-tw}}^{\left(
j\right) }+\hat{H}_{\text{cav-cav}}^{\left( j\right) }+\hat{H}_{\text{tw-cav}}^{\left(
j\right) } + \hat{H}_{\text{tw-Gtw}}^{\left( j\right) } + \hat{H}_{\text{cav-Gcav}}^{\left( j\right) }
 + \hat{H}_{\text{tw-Gcav}}^{\left( j\right) } + \hat{H}_{\text{cav-Gtw}}^{\left( j\right) }\right], \label{8}
\end{align}%
where the subscripts of these Hamiltonian parts denote those fields involved in these interactions. In what follows, we derive the detailed form of these terms. Since the particles are trapped at the focus of the optical
tweezers, the electric field at the particle position can be approximated as the potential at the focus (i.e., the Taylor expansion of the electric field around the focus). Therefore, the interaction term created by the tweezer field is given by
\begin{eqnarray}
\hat{H}_{\text{tw-tw}}^{\left( j\right) } &=&-\alpha \mathbf{E}%
_{\text{tw}j}^{2}( \mathbf{\hat{R}}_{j},t) /2 \notag \\
&=&-\frac{1}{2}\alpha \left\{ \frac{1}{2}\left[ \mathbf{E}%
_{\text{tw}j}( \mathbf{\hat{R}}_{j}) e^{-i\omega _{\text{tw}}t}+\mathbf{E}%
_{\text{tw}j}^{\ast }( \mathbf{\hat{R}}_{j}) e^{i\omega _{\text{tw}}t}\right]
\right\} ^{2} \notag \\
&\approx &\frac{1}{2}\alpha \epsilon _{\text{tw}j}^{2}\frac{1%
}{2}\left[ \frac{2 \hat{x}_{j} ^{2}}{A_{x_{j}}^{2} W_{t}^{2} }+\frac{2 \hat{y}_{j}^{2}}{A_{y_{j}}^{2} W_{t}^{2} }+\frac{\hat{z}_{j}^{2}}{z_{R}^{2}}\right]  \notag \\
&= &\sum_{\mu=x,y,z}\frac{1}{2}m\omega _{j\mu}^{2}\hat{\mu}_{j}^{2},\label{11}
\end{eqnarray}%
where the trapping frequencies of the harmonic trapping potential of the NP exerted by the tweezer are given by
\begin{eqnarray}
(\omega _{jx},\omega _{jy},\omega _{jz}) =\sqrt{\frac{\alpha \epsilon _{\text{tw}j}^{2}}{ 2m W_{t}^{2}}}\left( \sqrt{2}A_{x_{j}}^{-1},\sqrt{2}A_{y_{j}}^{-1}, \frac{\lambda _{\text{tw}}}{\pi W_{t}}\right) .
\end{eqnarray}%
We see from Eq.~(\ref{11}) that the interaction term describes the potential for a standard harmonic oscillator. Note that we make the rotating-wave approximation during the derivation of Eq.~(\ref{11}) by neglecting the rapidly oscillating terms with $\exp \left( \pm 2i\omega _{\text{tw}}t\right) $.

The second term in Eq.~(\ref{8}) is the squared term of the cavity field,
\begin{eqnarray}
\hat{H}_{\text{cav-cav}}^{\left( j\right) } &=&- \alpha \mathbf{E}%
_{\text{cav}}^{2}( \mathbf{\hat{R}}_{j})/2 \notag \\
&=& -\alpha \epsilon
_{\text{cav}}^{2}\cos ^{2}( k \hat{X}_{j}^{c}) \hat{a}^{\dag }\hat{a}  \notag \\
&=&-\alpha \epsilon
_{\text{cav}}^{2}\cos ^{2}[ k(\hat{X}_{j}\cos \theta +\hat{Y}_{j}\sin \theta)] \hat{a}^{\dag }\hat{a}  \notag \\
&\approx &\hbar \omega _{\text{sh}}^{\left( j\right) }\hat{a}^{\dag }\hat{a}+\hbar
g_{ax_{j}}\hat{a}^{\dag }\hat{a}\hat{x}_{j}+\hbar g_{ay_{j}}\hat{a}^{\dag }%
\hat{a}\hat{y}_{j} ,    \label{A4}
\end{eqnarray}%
which comprises of the radiation-pressure term and the frequency-shift terms
along the $X$- and $Y$-directions. The frequency shift is given by
\begin{eqnarray}
\omega _{\text{sh}}^{\left( j\right)
}=-\frac{\alpha}{\hbar} \epsilon _{\text{cav}}^{2}\cos ^{2}( kx_{j0}\cos \theta +ky_{j0}\sin \theta ) ,
\end{eqnarray}%
and the optomechanical coupling strengths along the $X$- and $Y$-directions are given by
\begin{subequations}
\begin{align}
g_{ax_{j}} &=\frac{\alpha}{\hbar} \epsilon _{\text{cav}}^{2}k\cos \theta \sin [ 2(
kx_{j0}\cos \theta +ky_{j0}\sin \theta ) ]  , \\
g_{ay_{j}} &=\frac{\alpha}{\hbar} \epsilon _{\text{cav}}^{2}k\sin \theta \sin [ 2(
kx_{j0}\cos \theta +ky_{j0}\sin \theta ) ]  .
\end{align}
\end{subequations}

In Eq.~(\ref{8}), the interaction term between the cavity field and the optical tweezer field is expressed as
\begin{eqnarray}
\hat{H}_{\text{tw-cav}}^{\left( j\right) } &=&-\alpha \mathbf{E}%
_{\text{tw}}^{\left( j\right) }( \mathbf{\hat{R}}_{j},t) \cdot \mathbf{E}%
_{\text{cav}}( \mathbf{\hat{R}}_{j}) \notag \\
&=&-\alpha \frac{1}{2}[ \mathbf{E}_{\text{tw}j}( \mathbf{\hat{R}}%
_{j}) e^{-i\omega _{\text{tw}}t}+\mathbf{E}_{\text{tw}j}^{\ast }( \mathbf{\hat{R%
}}_{j}) e^{i\omega _{\text{tw}}t}] \mathbf{e}_{\text{tw}}^{( j) }  \cdot \epsilon _{\text{cav}}\cos ( k\hat{X}_{j}\cos \theta
+k\hat{Y}_{j}\sin \theta -\phi )  ( \hat{a}%
e^{-i\omega _{\text{tw}}t}+\hat{a}^{\dag }e^{i\omega _{\text{tw}}t}) \mathbf{e}_{\text{cav}} \notag \\
&\approx& -\frac{1}{2}\alpha \epsilon _{\text{cav}} \epsilon_{\text{tw}}^{(
j) }[\cos ( kx_{j0}\cos \theta +ky_{j0}\sin \theta -\phi ) -k\cos \theta \sin ( kx_{j0}\cos \theta +ky_{j0}\sin \theta -\phi
)  \hat{x}_{j} \notag \\
&&-k\sin \theta \sin ( kx_{j0}\cos \theta +ky_{j0}\sin \theta -\phi
) \hat{y}_{j}] [ ( 1-ik_{\text{tw}}\hat{z}_{j}) \hat{a}^{\dag }+(
1+ik_{\text{tw}}\hat{z}_{j}) \hat{a}] \cos \theta \notag \\
&=&\hbar \Omega ^{\left( j\right) }( \hat{a}+\hat{a}^{\dag })
+\hbar g_{x_{j}}( \hat{a}+\hat{a}^{\dag }) \hat{x}_{j}  +\hbar g_{y_{j}}( \hat{a}+\hat{a}^{\dag }) \hat{y}_{j}+i\hbar
g_{z_{j}}( \hat{a}-\hat{a}^{\dag }) \hat{z}_{j},   \label{A7}
\end{eqnarray}%
where the driving amplitude of the cavity mode is given by
\begin{eqnarray}
\Omega ^{\left( j\right) } &=& -\frac{\alpha}{2\hbar} \epsilon _{\text{cav}}\epsilon
_{\text{tw}}^{\left( j\right) }\cos \theta \cos ( kx_{j0}\cos \theta
+ ky_{j0}\sin \theta )  ,
\end{eqnarray}%
and the coherent-scattering-mediated coupling strengths along the $X$-, $Y$-%
, and $Z$-directions are, respectively, given by
\begin{subequations}\label{18}
\begin{align}
g_{x_{j}} &=\frac{\alpha}{2\hbar} \epsilon _{\text{cav}}\epsilon _{\text{tw}}^{\left( j\right) }k\cos
^{2}\theta \sin (kx_{j0}\cos \theta +ky_{j0}\sin \theta) ,\\
g_{y_{j}} &=\frac{\alpha}{4\hbar} \epsilon _{\text{cav}}\epsilon _{\text{tw}}^{\left( j\right) }k\sin(
2\theta) \sin (kx_{j0}\cos \theta +ky_{j0}\sin \theta) ,\\
g_{z_{j}} &=-\frac{\alpha}{2\hbar} \epsilon _{\text{cav}}\epsilon _{\text{tw}}^{\left( j\right) }k_{\text{tw}}\cos
\theta \cos ( kx_{j0}\cos \theta +ky_{j0}\sin \theta ).
\end{align}
\end{subequations}

The remaining four terms in Eq.~(\ref{8}) describe the interaction between the incident field on the $j$th particle and the radiation field produced by the dipole of the $\bar{j}$th
particle at the position $\mathbf{\hat{R}}_{j}$, which can be written as%
\begin{subequations} \label{appendix}
\begin{align}
\hat{H}_{\text{tw-Gtw}}^{\left( j\right) }&=-\alpha \mathbf{E}%
_{\text{tw}}^{\left( j\right) }( \mathbf{\hat{R}}_{j},t) \cdot \mathbf{E}%
_{\text{Gtw}}^{\left( \bar{j}\right) }( \mathbf{\hat{R}}_{j},t) ,  \label{14a} \\
\hat{H}_{\text{cav-Gcav}}^{\left( j\right) }&=-\alpha \mathbf{E}%
_{\text{cav}}( \mathbf{\hat{R}}_{j}) \cdot \mathbf{E}_{\text{Gcav}}(
\mathbf{\hat{R}}_{j}) , \label{14b}  \\
\hat{H}_{\text{tw-Gcav}}^{\left( j\right) }&=-\alpha \mathbf{E}%
_{\text{tw}}^{\left( j\right) }( \mathbf{\hat{R}}_{j},t) \cdot \mathbf{E}%
_{\text{Gcav}}( \mathbf{\hat{R}}_{j}) ,   \label{14c} \\
\hat{H}_{\text{cav-Gtw}}^{\left( j\right) }&=-\alpha \mathbf{E}%
_{\text{cav}}( \mathbf{\hat{R}}_{j}) \cdot \mathbf{E}_{\text{Gtw}}^{\left( \bar{%
j}\right) }( \mathbf{\hat{R}}_{j},t) . \label{14d}
\end{align}
\end{subequations}%
We consider the interactions in the far-field regime $k_{0}R_{0}\gg 1,$ then the Green function can be approximated as
\begin{eqnarray}
\alpha \overleftrightarrow{\mathbf{G}}\left( \mathbf{R}_{0}\right)  &\approx
&e^{ik_{0}R_{0}}\eta _{f}( D/R_{0}^{3})  [( R_{0}^{2}-x^{2}) \mathbf{e}_{x}\mathbf{e}_{x}-xy\mathbf{e}_{x}\mathbf{e%
}_{y}  -xz\mathbf{e}_{x}\mathbf{e}_{z} -xy\mathbf{e}_{y}\mathbf{e}_{x} \nonumber \\
&& +( R_{0}^{2}-y^{2}) \mathbf{e}_{y}\mathbf{e}_{y} -yz\mathbf{e}_{y}\mathbf{e}_{z} -xz\mathbf{e}_{z}\mathbf{e}_{x} -yz\mathbf{e}_{z}\mathbf{e}_{y}+( R_{0}^{2}-z^{2}) \mathbf{e}%
_{z}\mathbf{e}_{z}],
\end{eqnarray}
where $\eta _{f}= \alpha k_{0}^{2}/4\pi \varepsilon _{0}D$ is the far-field constant.

In Eq.~(\ref{14a}), $\hat{H}_{\text{tw-Gtw}}^{\left(
j\right) }$ describes the transverse binding between the two NPs, and
it takes the form as%
\begin{eqnarray}
\hat{H}_{\text{tw-Gtw}}^{\left( j\right) } =\hbar R_{\alpha }\left( \hat{x}_{1}-%
\hat{x}_{2}\right) +\hbar R_{\beta }\left( \hat{y}_{1}-\hat{y}_{2}\right)
+\sum_{\mu=x,y,z}[ v_{\mu}( \hat{\mu}_{1}^{2}+\hat{\mu}_{2}^{2}) +%
\frac{1}{2}k_{\mu}\left( \hat{\mu}_{1}-\hat{\mu}_{2}\right) ^{2}]
+\sum_{j=1,2}\frac{1}{2}k_{xy}\hat{x}_{j}( \hat{y}_{j}-\hat{y}_{\bar{j}}) .\label{A1}
\end{eqnarray}%
Here, the displacement magnitudes are given by%
\begin{subequations}
\begin{align}
R_{\alpha} &= \alpha \eta_{f_{\text{tw}}}\mathcal{\epsilon}_{\text{tw}}^{(1)}\mathcal{\epsilon}_{\text{tw}}^{(2)}\bigl[ D^{-1}\cos (k_{\text{tw}}D) \cos^{3}\theta
           +k_{\text{tw}}\sin (k_{\text{tw}}D) \cos^{3}\theta
          -2D^{-1}\cos \theta \sin^{2}\theta \cos (k_{\text{tw}}D) \bigr]/\hbar , \\
R_{\beta} &= \alpha \eta_{f_{\text{tw}}}\mathcal{\epsilon}_{\text{tw}}^{(1)}\mathcal{\epsilon}_{\text{tw}}^{(2)}\bigl[ D^{-1}\cos (k_{\text{tw}}D) (2\sin^{3}\theta -2\sin \theta)
          +k_{\text{tw}}\sin (kD) \cos^{2}\theta \sin \theta
          +D^{-1}\cos^{2}\theta \sin \theta \cos (k_{\text{tw}}D) \bigr]/\hbar ,
\end{align}
\end{subequations}%
and $v_{\mu}$ ($\mu=x,y,z$) describes the frequency shift of the center-of-mass motion for the nanoparticles,%
\begin{subequations}
\begin{align}
v_{x} & = \alpha \eta_{f_{\text{tw}}}\mathcal{\epsilon}_{\text{tw}}^{(1)}\mathcal{\epsilon}_{\text{tw}}^{(2)}\cos^{2}\theta \cos(k_{\text{tw}}D) /(A_{x} W_{t})^{2}, \\
v_{y} &= \alpha \eta_{f_{\text{tw}}}\mathcal{\epsilon}_{\text{tw}}^{(1)}\mathcal{\epsilon}_{\text{tw}}^{(2)}\cos^{2}\theta \cos(k_{\text{tw}}D) /(A_{y} W_{t})^{2}, \\
v_{z} &= \alpha \eta_{f_{\text{tw}}}\mathcal{\epsilon}_{\text{tw}}^{(1)}\mathcal{\epsilon}_{\text{tw}}^{(2)}\cos^{2}\theta \cos(k_{\text{tw}}D) / (2z_{R}^{2}).
\end{align}
\end{subequations}%
In Eq.~(\ref{A1}), $k_{\mu}$ ($\mu=x,y,z$) represent the coupling strength between the particles mediated by coherent scattering, and $k_{xy}$ is the coupling coefficients associated with the $X$- and $Y$-direction motions,
\begin{subequations}
\begin{align}
k_{x} &= \alpha \eta_{f_{\text{tw}}}\mathcal{\epsilon}_{\text{tw}}^{(1)}\mathcal{\epsilon}_{\text{tw}}^{(2)}\bigl[ k_{\text{tw}}^{2}\cos(k_{\text{tw}}D) \cos^{4}\theta + D^{-2}\cos(k_{\text{tw}}D) (12\cos^{2}\theta \sin^{2}\theta - 2\sin^{2}\theta - 3\cos^{4}\theta + \cos^{2}\theta) \notag \\
       &\quad + D^{-1}k_{\text{tw}}\sin(k_{\text{tw}}D) (4\cos^{2}\theta \sin^{2}\theta - 3\cos^{4}\theta + \cos^{2}\theta) \bigr], \\
k_{y} &= \alpha \eta_{f_{\text{tw}}}\mathcal{\epsilon}_{\text{tw}}^{(1)}\mathcal{\epsilon}_{\text{tw}}^{(2)}\bigl[ D^{-2}\cos(k_{\text{tw}}D) (12\sin^{4}\theta - 14\sin^{2}\theta
- 3\cos^{2}\theta \sin^{2}\theta + \cos^{2}\theta + 2) \notag \\
  &\quad + k_{\text{tw}}D^{-1}\sin(k_{\text{tw}}D)
      (4\sin^{4}\theta - 4\sin^{2}\theta - 3\sin^{2}\theta \cos^{2}\theta + \cos^{2}\theta)+ k_{\text{tw}}^{2}\cos(k_{\text{tw}}D) \sin^{2}\theta \cos^{2}\theta \bigr], \\
k_{z} &= \alpha \eta_{f_{\text{tw}}}\mathcal{\epsilon}_{\text{tw}}^{(1)}\mathcal{\epsilon}_{\text{tw}}^{(2)}\bigl[ k_{\text{tw}}^{2}\cos(k_{\text{tw}}D) \cos^{2}\theta + k_{\text{tw}}D^{-1}
       \sin(k_{\text{tw}}D) \cos^{2}\theta \notag \\
        &\quad  + D^{-2}\cos(k_{\text{tw}}D) \cos^{2}\theta + 2D^{-2}\sin^{2}\theta \cos(k_{\text{tw}}D) \bigr], \\
k_{xy} &= \alpha \eta_{f_{\text{tw}}}\mathcal{\epsilon}_{\text{tw}}^{(1)}\mathcal{\epsilon}_{\text{tw}}^{(2)}\bigl[ D^{-2}\cos(k_{\text{tw}}D) (-5\cos\theta \sin^{3}\theta
+ 3\sin\theta \cos\theta - 3\cos^{3}\theta \sin\theta)  \notag \\
 &\quad + D^{-1}k_{\text{tw}}\sin(k_{\text{tw}}D)
        (4\cos\theta \sin^{3}\theta - 3\cos^{3}\theta \sin\theta - 2\sin\theta \cos\theta)+ k_{\text{tw}}^{2}\cos(k_{\text{tw}}D) \cos\theta \sin\theta \bigr].
\end{align}
\end{subequations}

The second cross term, namely the term in Eq.~(\ref{14b}), represents the longitudinal binding between the two nanoparticles. In terms of the involved fields, this interaction term can be obtained as%
\begin{eqnarray}
\hat{H}_{\text{cav-Gcav}} =-4\alpha \epsilon _{\text{cav}}^{2}\eta
_{f}\cos \left( kD\right) \cos ^{2}\left( kD/2\right) \hat{a}^{\dag }\hat{a}
+\hbar g_{\alpha }\left( \hat{x}_{1}-\hat{x}_{2}\right) ( \hat{a}%
^{\dag }\hat{a}+1/2) +\hbar g_{\beta }\left( \hat{y}_{1}-\hat{y}_{2}\right) ( \hat{a}%
^{\dag }\hat{a}+1/2) , \label{A16}
\end{eqnarray}%
where the optomechanical coupling strengths between the cavity field and the $X$- and $Y$-direction motions of the two particles are given by%
\begin{subequations}\label{A6}
\begin{align}
g_{\alpha} &= 4\alpha \epsilon_{\text{cav}}^{2}\eta_{f} \bigl\{ [ D^{-1}\cos(kD) + k\sin(kD) ] \cos\theta \cos^{2}(kD/2)
          + k\cos\theta \cos(kD) \sin(kD) / 2 \bigr\} / \hbar, \\
g_{\beta} &= 4\alpha \epsilon_{\text{cav}}^{2}\eta_{f} \bigl\{ [D^{-1}\cos(kD) + k\sin(kD)] \cos^{2}(kD/2)
          + k\cos(kD) \sin(kD) / 2 \bigr\} \sin\theta / \hbar .
\end{align}
\end{subequations}%

The third term, given by Eq.~(\ref{14c}), characterizes the interaction between the $j$th optical
tweezer field at $\mathbf{\hat{R}}_{j}$ and the radiation field of the $%
\bar{j}$th dipole induced by the cavity field. This term can be obtained with the related fields as follows,
\begin{eqnarray}
\hat{H}_{\text{tw-Gcav}} =\hbar \Omega _{\alpha }( \hat{a}%
^{\dag }+\hat{a}) +\sum_{j=1,2}\hbar g_{\alpha x_{j}}( \hat{a}%
^{\dag }+\hat{a}) \hat{x}_{j}
+\hbar g_{\alpha y_{j}}( \hat{a}^{\dag }+\hat{a}) \hat{y}%
_{j}+i\hbar g_{\alpha z_{j}}( \hat{a}-\hat{a}^{\dag }) \hat{z}%
_{j}. \label{A18}
\end{eqnarray}%
Here, the cavity displacement term $\Omega_{\alpha}$ and the coupling strength $g_{\alpha \mu_{j}}$ are given by%
\begin{subequations}\label{A8}
\begin{align}
\Omega_{\alpha} &= -\alpha \epsilon_{\text{cav}}\eta_{f}( \mathcal{\epsilon}_{\text{tw}}^{(1)}+\mathcal{\epsilon}_{\text{tw}}^{(2)}) \cos\theta \cos(kD) \cos(kD/2) /(2\hbar), \\
g_{\alpha x_{1}} &= \alpha \epsilon_{\text{cav}}\eta_{f}\bigl[ D^{-1}( \mathcal{\epsilon}_{\text{tw}}^{(1)}+\mathcal{\epsilon}_{\text{tw}}^{(2)}) \cos(kD) \cos(kD/2) \cos(2\theta)
 +(\mathcal{\epsilon}_{\text{tw}}^{(1)}+\mathcal{\epsilon}_{\text{tw}}^{(2)}) k\sin(kD) \cos(kD/2) \cos^{2}\theta \notag \\
                 &\quad +\mathcal{\epsilon}_{\text{tw}}^{(2)}\cos(kD) k\sin(kD/2) \cos^{2}\theta \bigr] /(2\hbar), \\
g_{\alpha x_{2}} &= -\alpha \epsilon_{\text{cav}}\eta_{f}\bigl[ D^{-1}( \mathcal{\epsilon}_{\text{tw}}^{(1)}+\mathcal{\epsilon}_{\text{tw}}^{(2)}) \cos(kD) \cos(kD/2) \cos(2\theta)
+(\mathcal{\epsilon}_{\text{tw}}^{(1)}+\mathcal{\epsilon}_{\text{tw}}^{(2)}) k\sin(kD) \cos(kD/2) \cos^{2}\theta \notag \\
                 &\quad +\mathcal{\epsilon}_{\text{tw}}^{(1)}\cos(kD) k\sin(kD/2) \cos^{2}\theta \bigr] /(2\hbar), \\
g_{\alpha y_{1}} &= \alpha \epsilon_{\text{cav}}\eta_{f}\bigl[ D^{-1}( \mathcal{\epsilon}_{\text{tw}}^{(1)}+\mathcal{\epsilon}_{\text{tw}}^{(2)}) \cos(kD) \cos(kD/2) \sin(2\theta)
  +(\mathcal{\epsilon}_{\text{tw}}^{(1)}+\mathcal{\epsilon}_{\text{tw}}^{(2)}) k\sin(kD) \cos(kD/2) \cos\theta \sin\theta \notag \\
                 &\quad +\mathcal{\epsilon}_{\text{tw}}^{(2)}\cos(kD) k\sin(kD/2) \cos\theta \sin\theta \bigr] /(2\hbar), \\
g_{\alpha y_{2}} &= -\alpha \epsilon_{\text{cav}}\eta_{f}\bigl[ D^{-1}( \mathcal{\epsilon}_{\text{tw}}^{(1)}+\mathcal{\epsilon}_{\text{tw}}^{(2)}) \cos(kD) \cos(kD/2) \sin(2\theta)
 +(\mathcal{\epsilon}_{\text{tw}}^{(1)}+\mathcal{\epsilon}_{\text{tw}}^{(2)}) k\sin(kD) \cos(kD/2) \cos\theta \sin\theta \notag \\
                 &\quad +\mathcal{\epsilon}_{\text{tw}}^{(1)}\cos(kD) k\sin(kD/2) \cos\theta \sin\theta \bigr] /(2\hbar), \\
g_{\alpha z_{j}} &= -\alpha \epsilon_{\text{cav}}\eta_{f}\mathcal{\epsilon}_{\text{tw}}^{(j)} k_{\text{tw}}\cos(kD) \cos(kD/2) \cos\theta /(2\hbar).
\end{align}
\end{subequations}

The fourth term, given by Eq.~(\ref{14d}), describes the interaction between the cavity field at the position $\mathbf{\hat{R}}_{j}$ and the radiation field produced by the
dipole induced by the $\bar{j}$th optical tweezer, and it is given by%
\begin{eqnarray}
\hat{H}_{\text{cav-Gtw}}  =\hbar ( \Omega _{\beta }\hat{a}%
+\Omega _{\beta }^{\ast }\hat{a}^{\dag }) +\sum_{j=1,2}\hbar (
g_{\beta x_{j}}\hat{a}+g_{\beta x_{j}}^{\ast }\hat{a}^{\dag })
\hat{x}_{j}
+\sum_{j=1,2}\hbar ( g_{\beta y_{j}}\hat{a}+g_{\beta
y_{j}}^{\ast }\hat{a}^{\dag }) \hat{y}_{j}
+\sum_{j=1,2}i\hbar ( g_{\beta z_{j}}\hat{a}-g_{\beta
z_{j}}^{\ast }\hat{a}^{\dag }) \hat{z}_{j},  \label{A20}
\end{eqnarray}%
where the cavity displacement amplitude $\Omega_{\beta}$ and the coupling strength $g_{\beta \mu_{j}}$ between the $Y$- and $Z$-direction motions of the particles are denoted as%
\begin{subequations}\label{A10}
\begin{align}
\Omega_{\beta} &= -\alpha \epsilon_{\text{cav}}\eta_{f_{\text{tw}}}( \mathcal{\epsilon}_{\text{tw}}^{\left( 1\right) }+\mathcal{\epsilon}_{\text{tw}}^{\left( 2\right)}) \cos \theta \cos \left( kD/2\right) e^{-ik_{\text{tw}}D}/2\hbar , \\
g_{\beta x_{1}} &= \alpha \epsilon_{\text{cav}}\eta_{f_{\text{tw}}}[D^{-1}(\mathcal{\epsilon}_{\text{tw}}^{\left( 1\right) }+\mathcal{\epsilon}_{\text{tw}}^{\left( 2\right) }) \cos \left( 2\theta \right) \cos \left( kD/2\right) +ik_{\text{tw}}( \mathcal{\epsilon}_{\text{tw}}^{\left( 1\right) }+\mathcal{\epsilon}_{\text{tw}}^{\left( 2\right) }) \cos ^{2}\theta \cos \left( kD/2\right) \notag \\
                 &\quad +\mathcal{\epsilon}_{\text{tw}}^{\left( 2\right) }k\cos ^{2}\theta \sin \left( kD/2\right) ]e^{-ik_{\text{tw}}D}/2\hbar , \\
g_{\beta x_{2}} &= -\alpha \epsilon_{\text{cav}}\eta_{f_{\text{tw}}}[D^{-1}(\mathcal{\epsilon}_{\text{tw}}^{\left( 1\right) }+\mathcal{\epsilon}_{\text{tw}}^{\left( 2\right) }) \cos \left( 2\theta \right) \cos \left( kD/2\right)
+ik_{\text{tw}}( \mathcal{\epsilon}_{\text{tw}}^{\left( 1\right) }+\mathcal{\epsilon}_{\text{tw}}^{\left( 2\right) }) \cos ^{2}\theta \cos \left( kD/2\right) \notag \\
                 &\quad +\mathcal{\epsilon}_{\text{tw}}^{\left( 1\right) }k\cos ^{2}\theta \sin \left( kD/2\right) ]e^{-ik_{\text{tw}}D}/2\hbar , \\
g_{\beta y_{1}} &= \alpha \epsilon_{\text{cav}}\eta_{f_{\text{tw}}}[2D^{-1}(\mathcal{\epsilon}_{\text{tw}}^{\left( 1\right) }+\mathcal{\epsilon}_{\text{tw}}^{\left( 2\right) }) \sin \theta \cos \theta \cos \left( kD/2\right) +ik_{\text{tw}}( \mathcal{\epsilon}_{\text{tw}}^{\left( 1\right) }+\mathcal{\epsilon}_{\text{tw}}^{\left( 2\right) }) \sin \theta \cos \theta \cos \left( kD/2\right) \notag \\
                 &\quad +\mathcal{\epsilon}_{\text{tw}}^{\left( 2\right) }k\sin \theta \cos \theta \sin \left( kD/2\right) ]e^{-ik_{\text{tw}}D}/2\hbar , \\
g_{\beta y_{2}} &= -\alpha \epsilon_{\text{cav}}\eta_{f_{\text{tw}}}[2D^{-1}(\mathcal{\epsilon}_{\text{tw}}^{\left( 1\right) }+\mathcal{\epsilon}_{\text{tw}}^{\left( 2\right) }) \sin \theta \cos \theta \cos \left( kD/2\right)  +ik_{\text{tw}}\left( \mathcal{\epsilon}_{\text{tw}}^{\left( 1\right) }+\mathcal{\epsilon}_{\text{tw}}^{\left( 2\right) }\right) \sin \theta \cos \theta \cos \left( kD/2\right) \notag \\
                 &\quad +\mathcal{\epsilon}_{\text{tw}}^{\left( 1\right) }k\sin \theta \cos \theta \sin \left( kD/2\right) ]e^{-ik_{\text{tw}}D}/2\hbar , \\
g_{\beta z_{j}} &= -\alpha \epsilon_{\text{cav}}\eta_{f}\mathcal{\epsilon}_{\text{tw}}^{\left( j\right) }k_{\text{tw}}\cos \theta \cos \left( kD/2\right) e^{-ik_{\text{tw}}D}/2\hbar .
\end{align}
\end{subequations}
By substituting Eqs.~(\ref{11}), ~(\ref{A4}), ~(\ref{A7}), ~(\ref{A1}), ~(\ref{A16}), ~(\ref{A18}), and ~(\ref{A20}) into Eq.~(\ref{8}), we can obtain the interaction Hamiltonian in Eq.~(\ref{appendix}).

\end{widetext}

\end{document}